\documentclass[reprint,pdflatex,twocolumn,superscriptaddress]{revtex4-2}
\usepackage[whole]{bxcjkjatype}

\usepackage{lineno}
\usepackage{amsmath}
\usepackage{etoolbox} 
\newcommand*\linenomathpatch[1]{%
  \cspreto{#1}{\linenomath}%
  \cspreto{#1*}{\linenomath}%
  \csappto{end#1}{\endlinenomath}%
  \csappto{end#1*}{\endlinenomath}%
}
\linenomathpatch{equation}
\linenomathpatch{gather}
\linenomathpatch{multline}
\linenomathpatch{align}
\linenomathpatch{alignat}
\linenomathpatch{flalign}

\RequirePackage{xcolor}
\RequirePackage{tikz}
\definecolor{myblack}{HTML}{4D4D4D}
\definecolor{mydarkgray}{HTML}{333333}
\definecolor{mygray}{HTML}{717171}
\definecolor{mywhite}{HTML}{F5F5F5}
\definecolor{myred}{HTML}{D04255}
\definecolor{myblue}{HTML}{4B75B9}
\definecolor{mygreen}{HTML}{00AF98}
\definecolor{myorange}{HTML}{EDAE00}
\definecolor{mypurple}{HTML}{C23685}
\definecolor{myviolet}{HTML}{9571CA}

\usepackage{tikz}
\usepackage{pxpgfmark}
\usetikzlibrary{arrows,positioning,plotmarks,external,patterns,angles,
decorations.pathmorphing,backgrounds,fit,shapes}
\usetikzlibrary{shapes.callouts}
\usepackage{amssymb}
\usepackage{bm}
\usepackage{txfonts}
\usepackage{newtxmath}
\usepackage{newtxtext}
\usepackage{physics}
\usepackage{float}
\usepackage{hyperref}
\hypersetup{
    colorlinks=true,
    citecolor=blue,
    linkcolor=blue,
    urlcolor=blue,
    pdfborder={0 0 0},
}
\usepackage{algorithm}
\usepackage{algpseudocode}
\usepackage{mathtools}
\usepackage{braket}
\usepackage{pgfplots}
\pgfplotsset{compat=1.12}

\makeatletter
\renewcommand*\env@matrix[1][*\c@MaxMatrixCols c]{%
  \hskip -\arraycolsep
  \let\@ifnextchar\new@ifnextchar
  \array{#1}}
  \makeatother

\usepackage{booktabs}

\newcommand{\mebius}{\hspace{0pt}_{\scalebox{1.5}{$\bullet$}} }
\newcommand\eq[1]{Eq.~(#1)}
\newcommand\eqs{Eqs.~}
\newcommand\headeq[1]{Equation~(#1)}

\newcommand\fig[1]{Fig.~#1}
\newcommand\figs[1]{Figs.~#1}
\newcommand\headfig[1]{Figure~#1}
\newcommand\headfigs[1]{Figures~#1}
\newcommand\sect[1]{Sec.~#1}

\newcommand\headapp[1]{Appendix~#1}

\newcommand{\fop}{\bm{{\mathrm{f}^\mathrm{o}}' }}
\newcommand{\fo}{\bm{\mathrm{f}^\mathrm{o}} }
\newcommand{\fep}{\bm{{\mathrm{f}^\mathrm{e}}'} }

\newcommand{\appsection}[1]{\section{\uppercase{#1}}}

\usepackage{ulem}
\usepackage{cancel}

\begin{document}

\title{Odd-frequency pairing in a nonunitary $p$-wave superconductor with multiple Majorana fermions}

\author{Daijiro Takagi}
\affiliation{Department of Applied Physics, Nagoya University, Nagoya 464-8603, Japan}

\author{Maria Teresa Mercaldo}
\affiliation{Dipartimento di Fisica ``E. R. Caianiello,'' Universit\'{a} di Salerno, IT-84084 Fisciano (SA), Italy}

\author{Yukio Tanaka}
\affiliation{Department of Applied Physics, Nagoya University, Nagoya 464-8603, Japan}

\author{Mario Cuoco}
\affiliation{Dipartimento di Fisica ``E. R. Caianiello,'' Universit\'{a} di Salerno, IT-84084 Fisciano (SA), Italy}
\affiliation{CNR-SPIN, IT-84084 Fisciano (SA), Italy}

\begin{abstract}
Odd-frequency Cooper pairs are gathering attention for the convenience of investigating the edge state of topological superconductors including Majorana fermions.
Although a spinless $p$-wave superconductor has only one Majorana fermion in a topological phase, the system with magnetic fields can reach the topological phases with multiple Majorana fermions.
To distinguish these multiple Majorana fermion phases, we correlate the energy spectrum with the odd-frequency pair amplitude as increasing the system size.
The system size dependence tells us three pieces of information: the parity of the number of the Majorana fermions at the edge, the number of low-energy modes corresponding to the Majorana fermions with different localization lengths, and the fingerprints of the Majorana fermions.
Also, we present the spatial dependence of the odd-frequency $\bm{\mathrm{f}}$ vector that is created from odd-frequency pair amplitude and the spin structure of odd-frequency Cooper pairs. 
We find that the odd-frequency $\bm{\mathrm{f}}$ vector is fixed in the same direction in any topological phase.
Also, we show that the spin state of odd-frequency Cooper pairs tends to be oriented toward the direction of the magnetic fields.
Our results highlight the odd-frequency Cooper pairs can be a good indicator for the detection of the multiple Majorana fermions and the distinction among the topological phases.
\end{abstract}

\maketitle

\section{Introduction}
\label{sect:intro}
An odd-frequency Cooper pair, proposed by Berezinskii, is a special couple of electrons~\cite{berezinskii1974new}.
Conventionally, the amplitude of Cooper pairs that constitute superconductivity shows an even-frequency dependence. 
Under the Fermi--Dirac statistics, which two electrons forming a Cooper pair satisfy, even-frequency spin-singlet even-parity and even-frequency spin-triplet odd-parity pairings can exist.
This is because the pair amplitude must be antisymmetric for the exchange of spin and position between two electrons.
Relaxing the restriction that the pair amplitude is even for frequency enables odd-frequency spin-triplet even-parity and odd-frequency spin-singlet odd-parity pairings to exist~\cite{berezinskii1974new,balatsky1992new}. 
Several works, however, theoretically indicate the thermodynamic instability~\cite{fominov2015odd,matsubara2021generation,coleman1995three,heid1995thermodynamic} of the odd-frequency Cooper pairs in the bulk of a homogeneous single-band superconductor~\cite{berezinskii1974new,balatsky1992new,kirkpatrick1991disorder,solenov2009thermodynamical,kusunose2011puzzle}
\footnote{
The odd-frequency pairing states in the bulk of a homogeneous system exhibit the paramagnetic Meissner effect contrary to the definition of the superconductivity~\cite{coleman1995three,heid1995thermodynamic}. 
On the other hand, by assuming a nonHermitian odd-frequency gap function, the diamagnetic Meissner effect emerges~\cite{kirkpatrick1991disorder,solenov2009thermodynamical,kusunose2011puzzle}. 
However, in this situation, the mixture between paramagnetic and diamagnetic odd-frequency pairing induces unphysical results: (i) the imaginary Josephson current~\cite{fominov2015odd} and (ii) the imaginary spin current~\cite{matsubara2021generation}.
}.
For this reason, the odd-frequency Cooper pairs being secondarily generated from an even-frequency superconductor are more promising~\cite{bergeret2005odd,tanaka2012symmetry,linder2019odd,cayao2020odd}.

Nonuniform systems~\cite{tanaka2007anomalous,tanaka2012symmetry,tanaka2007odd,eschrig2007symmetries} or systems with magnetic fields~\cite{bergeret2001long,bergeret2005odd,buzdin2005proximity,eschrig2015spin,golubov2004current,linder2015superconducting} induce the odd-frequency Cooper pairs.
Examples of the nonuniformity include edges (surfaces) and junctions.
Fascinating phenomena, caused by the odd-frequency Cooper pairs, are a long-range proximity effect in an $s$-wave superconductor / ferromagnet junction~\cite{bergeret2001long,bergeret2005odd,buzdin2005proximity,eschrig2015spin,golubov2004current,linder2015superconducting} and an anomalous proximity effect in a spin-triplet superconductor / diffusive normal metal junction~\cite{tanaka2007theory,tanaka2004anomalous,tanaka2005theory,asano2006anomalous,ikegaya2016quantization,takagi2020odd}.
In particular, the anomalous proximity effect is closely related to Majorana fermions and topological superconductors.

Majorana fermions are quasiparticles with no distinction between creation and annihilation.
They are expected to be applied as qubits because of their nonAbelian statistics~\cite{kitaev2003fault,nayak2008non}.
The Majorana fermions appear as zero-energy states at the edge of a topological superconductor~\cite{mourik2012signatures}.
The simplest model for representing Majorana fermions is the Kitaev chain: a spinless $p$-wave superconductor~\cite{kitaev2001unpaired}.
At each edge of the Kitaev chain, the number of Majorana fermions, corresponding to the winding number, is one.
Considering higher-order hopping and magnetic fields in a $p$-wave superconductor with spin degrees of freedom allows us to access topological phases with multiple Majorana fermions~\cite{mercaldo2016magnetic,mercaldo2018magnetic,sakurai2020nodal,mercaldo2019magnetoelectrically,sakurai2020nodal}.

There is a one-to-one correspondence between the Majorana fermions and the odd-frequency Cooper pairs~\cite{cayao2020odd,asano2013majorana,tamura2019odd,takagi2020odd,cayao2017odd,tsintzis2019odd,tamura2020bulk}.
Specifically, in the low-frequency limit, the amplitude of the normal Green's function (corresponding to the Majorana fermions) is equal to that of the anomalous Green's function (the odd-frequency pair amplitude)~\cite{asano2013majorana,tamura2019odd,takagi2020odd}.
The odd-frequency pair amplitude, accompanying the Majorana fermions, has a $1/\omega$ dependence, where $\omega$ is frequency~\cite{tamura2019odd}.
In addition, the sum of the odd-frequency pair amplitude is related to the winding number that is extended to a finite frequency in a limited semi-infinite system with chiral symmetry: spectral bulk-boundary correspondence~\cite{tamura2019odd,daido2019chirality}.

This paper aims to distinguish the number of Majorana fermions in terms of the odd-frequency Cooper pairs and to elucidate the spatial dependence and spin structure of the odd-frequency pair amplitude in the $p$-wave superconductor with the multiple Majorana fermions.
In the semi-infinite system with the multiple Majorana fermions, it is not easy to distinguish the topological phases with the different topological numbers (winding number) by focusing on the local density of states at the edge.
This is because these Majorana fermions appear as the same zero-energy states.
Therefore, we have concentrated on the odd-frequency Cooper pairs that have been useful for understanding the topological properties and transport phenomena of the Kitaev chain~\cite{takagi2020odd,mishra2021impact}.  

\begin{figure*}[htbp]
    \centering
    \includegraphics[width=0.99\textwidth]{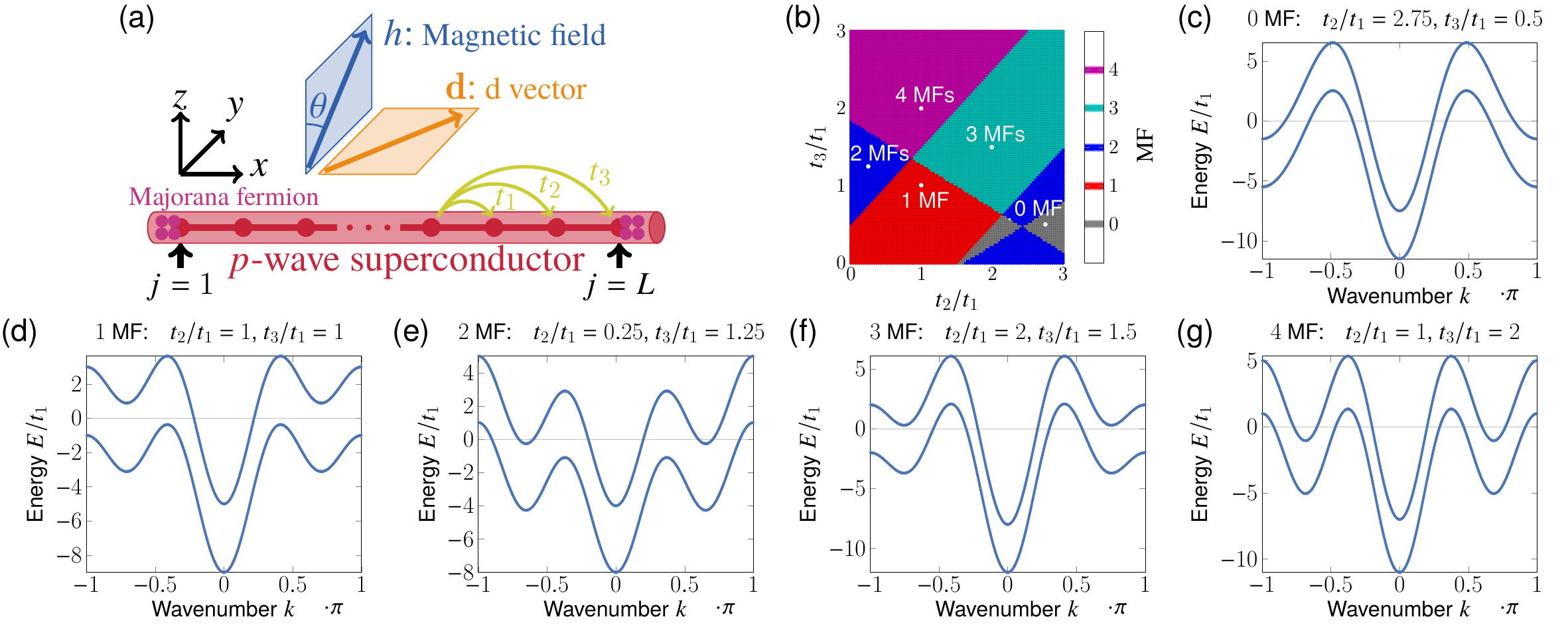}
    \caption{
        (a) A sketch of a $p$-wave superconductor (red tube) with magnetic fields.
    The superconductor lies in $x$ direction.
    The magnetic fields are applied in the $y$-$z$ plane.
    We set the $\bm{\mathrm{d}}$ vector in $x$-$y$ plane. 
    Hopping is considered up to the third nearest neighbor sites: ($t_1$, $t_2$, $t_3$).
    $j$: site, $L$: the number of sites in the system.
    (b) A topological phase diagram in the $p$-wave superconductor with the magnetic fields as a function of $t_2$ and $t_3$:
    $\Delta_{\uparrow\uparrow}=0.18t_1$, $\Delta_{\downarrow\downarrow}=1.8t_1$.
    The topological phases with the multiple Majorana fermions (MFs) ($w>2$) are obtained for $t_2$, $t_3 > t_1$.
    (c)--(g) Energy bands of normal states ($\Delta_{\uparrow\uparrow} = \Delta_{\downarrow\downarrow} = 0$) in $\bm{\mathrm{k}}$ space. (c) 0 MF, (d) 1 MF, (e) 2 MF, (f) 3 MF, (g) 4 MF phases, where MF means the Majorana fermion.
    Parameters in (b)--(g): $\theta = \pi/4$, $h/t_1=2$, $\mu/t_1=1$.
    \label{fig:system-phase-energy-normal}
    }
\end{figure*}

First, we investigate the system size dependence of the odd-frequency pair amplitude with and without magnetic fields that break chiral symmetry.
The system size dependence provides three pieces of information: the parity of the number of the Majorana fermions at the edge, the number of low-energy modes corresponding to the Majorana fermions with different localization lengths, and the fingerprints of the Majorana fermions.
Also, we present the spatial dependence of the odd-frequency $\bm{\mathrm{f}}$ vector representing odd-frequency pair amplitude and the spin structure of odd-frequency Cooper pairs. 
Here, the vector is defined by imitating the $\bm{\mathrm{d}}$ vector that characterizes spin-triplet superconductivity.
We find the following two results.
The vector is fixed in the same direction in any topological phase.
The spin direction of odd-frequency Cooper pairs tends to point toward the direction of the magnetic fields.
Our findings show that odd-frequency Cooper pairs can be an indicator for the detection of multiple Majorana fermions and the distinction among different topological phases.

Our paper is organized as follows. In \sect{\ref{sect:model}}, we introduce the model of a $p$-wave superconductor with multiple Majorana fermions. 
In \sect{\ref{sect:method}}, we explain how to calculate odd-frequency pair amplitude by using Green's function.
Then, we provide the system size dependence of low energy modes corresponding to Majorana fermions and odd-frequency pair amplitude in \sect{\ref{sect:odd-freq}}.
In addition, we present the spatial dependence of the odd-frequency $\bm{\mathrm{f}}$ vector representing odd-frequency pair amplitude and the spin structure of odd-frequency Cooper pairs in \sect{\ref{sect:spatial}}.
In \sect{\ref{sect:discussion}}, we discuss a setup for detecting evidence of the multiple Majorana fermions.
Finally, we conclude our work in \sect{\ref{sect:conclusion}}.

\section{Model}
\label{sect:model}
As the model for investigating the multiple Majorana phases, we introduce a $p$-wave superconductor with magnetic fields~\cite{mercaldo2016magnetic} shown in \fig{\ref{fig:system-phase-energy-normal}(a)}.

The Hamiltonian of the system on the lattice is written as 
\begin{align}
    \label{eq:hamiltonian-real}
    \mathcal{H} &= \mathcal{H}_\mathrm{N} + \mathcal{H}_{\bm{\mathrm{h} }} + \mathcal{H}_\mathrm{S},
\end{align}
\begin{align}
    \mathcal{H}_\mathrm{N} &= - \mu \sum_{j,\sigma} \left(c_{j,\sigma}^\dagger c_{j,\sigma}\right) 
                           - \sum_{j,\sigma,\nu} t_\nu \left(c_{j,\sigma}^\dagger c_{j+\nu,\sigma}+c_{j+\nu,\sigma}^\dagger c_{j,\sigma}\right)\nonumber,\\
                           \mathcal{H}_{\bm{\mathrm{h} }} &= -\sum_{j,\sigma,\sigma^\prime} \left(\bm{\mathrm{h}}\cdot\bm{\mathrm{\sigma}}\right)_{\sigma,\sigma^\prime} c_{j,\sigma}^\dagger c_{j,\sigma^\prime},\nonumber\\
    \mathcal{H}_\mathrm{S} &= \sum_{j,\sigma} \left[\Delta_{\sigma\sigma}c_{j,\sigma}^\dagger c_{j+1,\sigma}^\dagger+\mathrm{H.c.}\right].
\end{align}
where $c_{j,\sigma}$, $\mu$, $t_{\nu}$, $\bm{\mathrm{h}}=(h_x,h_y,h_z)=(0,h\sin\theta,h\cos\theta)$, and $\Delta_{\sigma,\sigma}$ ($\in \mathbb{R}$, assume spin-polarized case: the component $d_z=0$ of $\bm{\mathrm{d}}$ vector) denote the annihilation operator with a spin $\sigma$ at a site $j\in[1,L]$, chemical potential, the $\nu$-th ($\nu=1,2,3$) nearest-neighbor hopping, Zeeman potential, and pair potential, respectively~\cite{mercaldo2016magnetic}.
$\mathrm{H.c.}$ means Hermitian conjugate and $\bm{\mathrm{\sigma}}$ are Pauli matrices.

The Hamiltonian of the system with $L$ sites can be expressed in the block matrix form as
\begin{align}
    H = 
    \begin{bmatrix}
        \hat{u} & \hat{t}_1 & \hat{t}_2 & \hat{t}_3 & O &\cdots\\
        \hat{t}_1^\dagger & \hat{u} & \hat{t}_1 & \hat{t}_2 & \hat{t}_3&\\
        \hat{t}_2^\dagger & \hat{t}_1^\dagger & \hat{u} & \hat{t}_1 & \hat{t}_2 &\\
        \hat{t}_3^\dagger & \hat{t}_2^\dagger & \hat{t}_1^\dagger & \hat{u} & \hat{t}_1 &\\
        O & \hat{t}_3^\dagger & \hat{t}_2 & \hat{t}_1^\dagger & \hat{u} &\\
        \vdots &&&&&\ddots\\
    \end{bmatrix},
    \label{eq:hamiltonian-real-matrix}
\end{align}
where $\mathcal{H}=(1/2)[C_1^\dagger,C_2^\dagger,\ldots,C_L^\dagger]H[C_1,C_2,\ldots,C_L]^\mathsf{T}-\mu L$ with $C_j=[c_{j\uparrow},c_{j\downarrow},c_{j\uparrow}^\dagger,c_{j\downarrow}^\dagger]$,
\begin{align}
    \hat{u} &= 
    \begin{bmatrix}
        -\mu-h_z&ih_y&0&0\\
        -ih_y&-\mu+h_z&0&0\\
        0&0&\mu+h_z&ih_y\\
        0&0&-ih_y&\mu-h_z
    \end{bmatrix},\nonumber\\
    \hat{t}_1 &= 
    \begin{bmatrix}
        -t_1&0&\Delta_{\uparrow\uparrow}&0\\
        0&-t_1&0&\Delta_{\downarrow\downarrow}\\
        -\Delta_{\uparrow\uparrow}&0&t_1&0\\
        0&-\Delta_{\downarrow\downarrow}&0&t_1
    \end{bmatrix},\nonumber\\
    \hat{t}_2 &= 
    \begin{bmatrix}
        -t_2&0&0&0\\
        0&-t_2&0&0\\
        0&0&t_2&0\\
        0&0&0&t_2
    \end{bmatrix},\quad
    \hat{t}_3 = 
    \begin{bmatrix}
        -t_3&0&0&0\\
        0&-t_3&0&0\\
        0&0&t_3&0\\
        0&0&0&t_3
    \end{bmatrix}.
    \label{eq:hamiltonian-real-five}
\end{align}

By applying Fourier transformation to \eq{\ref{eq:hamiltonian-real}}, we get the Hamiltonian in $\bm{\mathrm{k}}$ space as
\begin{align}
    \label{eq:hamiltonian-k-one}
    \mathcal{H} (k) &= \sum_{k,\sigma}\varepsilon(k) c_{k,\sigma}^\dagger c_{k,\sigma}
    -\sum_{k,\sigma,\sigma^\prime} \left(\bm{\mathrm{h}}\cdot\bm{\mathrm{\sigma}}\right)_{\sigma,\sigma^\prime} c_{k,\sigma}^\dagger c_{k,\sigma^\prime}\nonumber\\
     &+ \sum_{k,\sigma} [i\Delta_{\sigma\sigma} \sin k c_{k,\sigma}^\dagger c_{-k,\sigma}^\dagger+\mathrm{H.c.}],  
\end{align}
where $k$ is the wavenumber and $\varepsilon(k)=-\mu-2t_1\cos k-2t_2\cos 2k -2t_3\cos 3k$.
\headeq{\ref{eq:hamiltonian-k-one}} is rewritten in matrix representation: 
\begin{align}
    \label{eq:hamiltonian-k-two}
    H(k) &= 
    \begin{bmatrix}
        \varepsilon(k) -h_z                      & ih_y                                   & 2i\Delta_{\uparrow\uparrow} \sin k & 0\\
        -ih_y                                & \varepsilon(k) +h_z                          & 0                                  & 2i\Delta_{\downarrow\downarrow} \sin k\\
        -2i\Delta_{\uparrow\uparrow} \sin k & 0                                            & -\varepsilon(k) +h_z               & ih_y &\\
        0                                        & -2i\Delta_{\downarrow\downarrow} \sin k & -ih_y                           & -\varepsilon(k) -h_z& \\
    \end{bmatrix},
\end{align}
where $\mathcal{H}(k) = (1/2) \sum_{k}C_k^\dagger H(k) C_k+\mbox{const.}$ with the base $C_k = [c_{k,\uparrow}, c_{k,\downarrow}, c_{-k,\uparrow}^\dagger, c_{-k,\downarrow}^\dagger]^\mathsf{T}$.

The winding number, telling us the number of the Majorana fermions at the edge of the system, is calculated from the Hamiltonian in $\bm{\mathrm{k}}$ space shown in \eq{\ref{eq:hamiltonian-k-two}}:

\begin{align}
    \label{eq:winding}
    w &= \frac{-1}{4\pi i} \int_{-\pi}^{\pi} \mathrm{Tr} \left[\Gamma H^{-1}(k) \partial_k H(k)\right] \mathrm{d}k\nonumber\\
      &= 
    \begin{cases}
        0 & (\mbox{trivial})\\
        1,2,3,4 & (\mbox{topological}).
    \end{cases}
\end{align}
$\Gamma=\tau_x\sigma_z$, defined for $h_x=0$, means the chiral operator satisfying $\{\Gamma,H(k)\}=0$.
In this system, the topological phase with $w>2$ can be achieved by controlling the parameters~\cite{mercaldo2016magnetic} as shown in \fig{\ref{fig:system-phase-energy-normal}(b)}.
The $\bm{\mathrm{d}}$ vector ($\bm{\mathrm{d}}=[(\Delta_{\downarrow\downarrow}-\Delta_{\uparrow\uparrow})/2, (\Delta_{\uparrow\uparrow}+\Delta_{\downarrow\downarrow})/2, \Delta_{\uparrow\downarrow}]$), characterizing the spin-triplet pairing, with more than one component and long-distance hoppings ($t_2$ and $t_3$) are necessary conditions to get 3 and 4 MF (Majorana fermion) phases.

The importance of long-distance hopping can be understood from the energy dispersion of the normal states in $k$ space. 
The energy dispersion, shown in \fig{\ref{fig:system-phase-energy-normal}(c)--\ref{fig:system-phase-energy-normal}(g)}, is calculated as $E=\varepsilon(k)\pm h$ by the diagonalization of the upper left $2\times2$ block matrix in Hamiltonian [\eq{\ref{eq:hamiltonian-k-two}}].
The number of Fermi points (intersections of energy bands and $E=0$) is more in the 2, 3, and 4 MF phases [\figs{\ref{fig:system-phase-energy-normal}(e)--\ref{fig:system-phase-energy-normal}}(g)] than in the 1 MF phase [\fig{\ref{fig:system-phase-energy-normal}(d)}], with the exception in the 0 MF phase [\fig{\ref{fig:system-phase-energy-normal}(c)}].
To have several Fermi points, $\varepsilon(k)$ needs to contain high-frequency components such as $\cos 2k$ and $\cos 3k$.
Higher-order hoppings $t_2$ and $t_3$ are a prerequisite for accessing the 3 MF and 4 MF phases in this system [In the strict sense, higher-order hopping promotes the sign changes in the real and imaginary parts of \eq{\ref{eq:winding-two}}].

\section{Method}
\label{sect:method}
To relate the multiple Majorana fermions and the odd-frequency Cooper pairs, we calculate the ground state energy corresponding to the Majorana fermion and the odd-frequency pair amplitude. 
The ground state energy can be obtained from the diagonalization of the Hamiltonian in the real space $H$ in \eq{\ref{eq:hamiltonian-real-matrix}}. 
The odd-frequency pair amplitude is calculated by using the Matsubara Green's function. 
For the large system, we use the recursive Green's function method~\cite{umerski1997closed} to calculate the Green's function at the required site numerically.

The Green's function at the site $j$, $j'$ is defined as
\begin{align}
    \label{eq:matsubara}
    \hat{G}(i\omega,j,j') &= \left\{[i\omega I- H]^{-1}\right\}_{j,j'}\nonumber\\
                    &=
    \begin{bmatrix}
        G_{j,j'}(i\omega) & F_{j,j'}(i\omega)\\
        \tilde{F}_{j,j'}(i\omega) & \tilde{G}_{j,j'}(i\omega)
    \end{bmatrix}
\end{align}
where $I$ represents an identity matrix.
$G(\tilde{G})[F(\tilde{F})]$ means the normal Green's function in the electron-electron (hole-hole) space [the anomalous Green's function in the electron-hole (hole-electron) space] with the $4\times 4$ sizes, and $\omega$ is Matsubara frequency.
Each local Green's function can be decomposed into spin components $\uparrow$ and $\downarrow$.
For example, the anomalous Green's function is expressed as
\begin{align}
    F_{j,j'} = 
    \begin{bmatrix}
        F_{j,j',\uparrow,\uparrow}   & F_{j,j',\uparrow,\downarrow}\\
        F_{j,j',\downarrow,\uparrow} & F_{j,j',\downarrow,\downarrow}
    \end{bmatrix},
\end{align}
and $G$, $\tilde{G}$, and $\tilde{F}$ are described, similarly.

The Matsubara Green's function shown in \eq{\ref{eq:matsubara}} gives the odd- (even-) frequency spin-triplet $s$-wave ($p$-wave) pair amplitude:
\begin{align}
    \label{eq:odd-even-freq-vector}
    \bm{\mathrm{f}^\mathrm{o(e)}}&=[f^\mathrm{o(e)}_x,f^\mathrm{o(e)}_y,f^\mathrm{o(e)}_z]\nonumber\\
       &=\left[\frac{f_{\downarrow\downarrow}^\mathrm{o(e)}-f_{\uparrow\uparrow}^\mathrm{o(e)}}{2},\frac{f_{\uparrow\uparrow}^\mathrm{o(e)}+f_{\downarrow\downarrow}^\mathrm{o(e)}}{2i},f_{\uparrow\downarrow}^\mathrm{o(e)}\right],
\end{align}
\begin{align}
    \label{eq:odd-freq}
    f^\mathrm{o}_{\sigma\sigma'}(j) &= \frac{1}{2}\left[{F}_{j,j,\sigma,\sigma'}(i\omega)-{F}_{j,j,\sigma,\sigma'}(-i\omega)\right],
\end{align}
\begin{align}
    \label{eq:even-freq}
    f_{\sigma\sigma'}^\mathrm{e}(j) &= 
    \frac{1}{2}
    \left[\frac{F_{j,j+1,\sigma,\sigma'}(i\omega)-F_{j+1,j,\sigma,\sigma'}(i\omega)}{2}\right.\nonumber\\
    &\qquad\left.+\frac{F_{j,j+1,\sigma,\sigma'}(-i\omega)-F_{j+1,j,\sigma,\sigma'}(-i\omega)}{2}
    \right].
\end{align}
where the superscript $^\mathrm{o}$ and $^\mathrm{e}$ stand for the ``odd'' and ``even'' frequencies, respectively.
Note that we only focus on the odd- (even-) frequency spin-triplet $s$-wave ($p$-wave) pair amplitude although there are Cooper pairs with other symmetries.
Here, we define the odd- and even-frequency pair amplitude as the vector form $\bm{\mathrm{f}^{\mathrm{o(e)} }}$ to discuss the spin state of these Cooper pairs. 

To get the Matsubara Green's function in the large finite system efficiently, we use the recurrence relation in the recursive Green's function method~\cite{umerski1997closed}.
The recurrence relation increasing sites to the ``right'' given by
\begin{align}
    \label{eq:recurrence-right}
    g_\mathrm{L}^{(n)} &= (\Xi_\mathrm{L})^n \mebius g_\mathrm{L}^{(0)},\\
    \Xi_\mathrm{L} &= 
    \begin{bmatrix}
        a_\mathrm{L} & b_\mathrm{L}\\
        c_\mathrm{L} & d_\mathrm{L}
    \end{bmatrix},\quad
    g_\mathrm{L}^{(n)} = 
    \begin{bmatrix}
        G_{n-2,n-2}^{(n)}&G_{n-2,n-1}^{(n)}&G_{n-2,n}^{(n)}\\
        G_{n-1,n-2}^{(n)}&G_{n-1,n-1}^{(n)}&G_{n-1,n}^{(n)}\\
        G_{n,n-2}^{(n)}&G_{n,n-1}^{(n)}&G_{n,n}^{(n)}\\
    \end{bmatrix},\nonumber\\
    a_\mathrm{L} &= 
    \begin{bmatrix}
        O&\hat{I}&O\\
        O&O&\hat{I}\\
        O&O&O
    \end{bmatrix},\quad
    b_\mathrm{L} = 
    \begin{bmatrix}
        O&O&O\\
        O&O&O\\
        \hat{t}_3^{-1}&O&O
    \end{bmatrix},\nonumber\\
    c_\mathrm{L} &= 
    \begin{bmatrix}
        O&O&O\\
        O&O&O\\
        -\hat{t}_3^\dagger&-\hat{t}_2^\dagger&-\hat{t}_1^\dagger
    \end{bmatrix},\quad
    d_\mathrm{L} = 
    \begin{bmatrix}
        -\hat{t}_2\hat{t}_3^{-1}&\hat{I}&O\\
        -\hat{t}_1\hat{t}_3^{-1}&O&\hat{I}\\
        (i\omega-u)\hat{t}_3^{-1}&O&O
    \end{bmatrix},
\end{align}
where $G_{j,j'}^{(n)}$ means the local Green's function at the $j$,$j'$ site in the system with $n$ sites.
The operator $\mebius$ represents the left-hand M\"{o}bius transformation defined as 
\begin{align}
    \begin{bmatrix}
        A & B\\
        C & D
    \end{bmatrix}
    \mebius Y \equiv (AY + B) (CY+D)^{-1}
\end{align}
with the square matrix $A$, $B$, $C$, $D$, and $Y$ with the same size as each other.
Similarly, the recurrence relation increasing sites to the ``left'' is obtained by 
\begin{align}
    \label{eq:recurrence-left}
    g_\mathrm{R}^{(n)} &= (\Xi_\mathrm{R})^n \mebius g_\mathrm{R}^{(0)},\\
    \Xi_\mathrm{R} &= 
    \begin{bmatrix}
        a_\mathrm{R} & b_\mathrm{R}\\
        c_\mathrm{R} & d_\mathrm{R}
    \end{bmatrix},\quad
    g_\mathrm{R}^{(n)} = 
    \begin{bmatrix}
        G_{1,1}^{(n)}&G_{1,2}^{(n)}&G_{1,3}^{(n)}\\
        G_{2,1}^{(n)}&G_{2,2}^{(n)}&G_{2,3}^{(n)}\\
        G_{3,1}^{(n)}&G_{3,2}^{(n)}&G_{3,3}^{(n)}\\
    \end{bmatrix},\nonumber\\
    a_\mathrm{R} &= 
    \begin{bmatrix}
        0&0&0\\
        \hat{I}&0&0\\
        0&\hat{I}&0
    \end{bmatrix},\quad
    b_\mathrm{R} = 
    \begin{bmatrix}
        0&0&\hat{t}_3^{\dagger -1}\\
        0&0&0\\
        0&0&0
    \end{bmatrix},\nonumber\\
    \label{eq:recurrence-left-componet}
    c_\mathrm{R} &= 
    \begin{bmatrix}
        -\hat{t}_1&-\hat{t}_2&-\hat{t}_3\\
        0&0&0\\
        0&0&0
    \end{bmatrix},\quad
    d_\mathrm{R} = 
    \begin{bmatrix}
        0&0&(i\omega-u)\hat{t}_3^{\dagger -1}\\
        \hat{I}&0&-\hat{t}_1^\dagger\hat{t}_3^{\dagger -1}\\
        0&\hat{I}&-\hat{t}_2^\dagger\hat{t}_3^{\dagger -1}
    \end{bmatrix}.
\end{align}

The local Green's function after connecting $g_\mathrm{L}^{(m)}$ and $g_\mathrm{R}^{(n)}$ is written as follows:
\begin{align}
    \left[g_\mathrm{M}^{(m+n)}\right]_{11} &=
    \begin{bmatrix}
        G_{m-2,m-2}^{(m+n)} & G_{m-2,m-1}^{(m+n)} & G_{m-2,m}^{(m+n)}\\
        G_{m-1,m-2}^{(m+n)} & G_{m-1,m-1}^{(m+n)} & G_{m-1,m}^{(m+n)}\\
        G_{m,m-2}^{(m+n)} & G_{m,m-1}^{(m+n)} & G_{m,m}^{(m+n)}
    \end{bmatrix}.
    \label{eq:connect}
\end{align}
We can get the local Green's function at the required site in the finite system by combining \eqs(\ref{eq:recurrence-right}), (\ref{eq:recurrence-left}), and (\ref{eq:connect}).

\section{Odd-frequency Cooper pairs and multiple Majorana fermions}
\label{sect:odd-freq}
In this section, we would like to elucidate the property of the Majorana fermion. 
In the semi-infinite system with the multiple Majorana fermions, it is, however, not easy to distinguish the topological phases with the different topological numbers (winding numbers) by focusing on the local density of states at the edge.
This is because the energy states hosting Majorana fermions belong to the same zero energy. 
Therefore, we have investigated the system size dependence of the low energy spectra and that of the odd-frequency pair amplitude.
Moreover, we have added the perturbation breaking the chiral symmetry which removes the multiple Majorana fermions.

\begin{figure}[!htbp]
    \centering
    \includegraphics[width=0.49\textwidth]{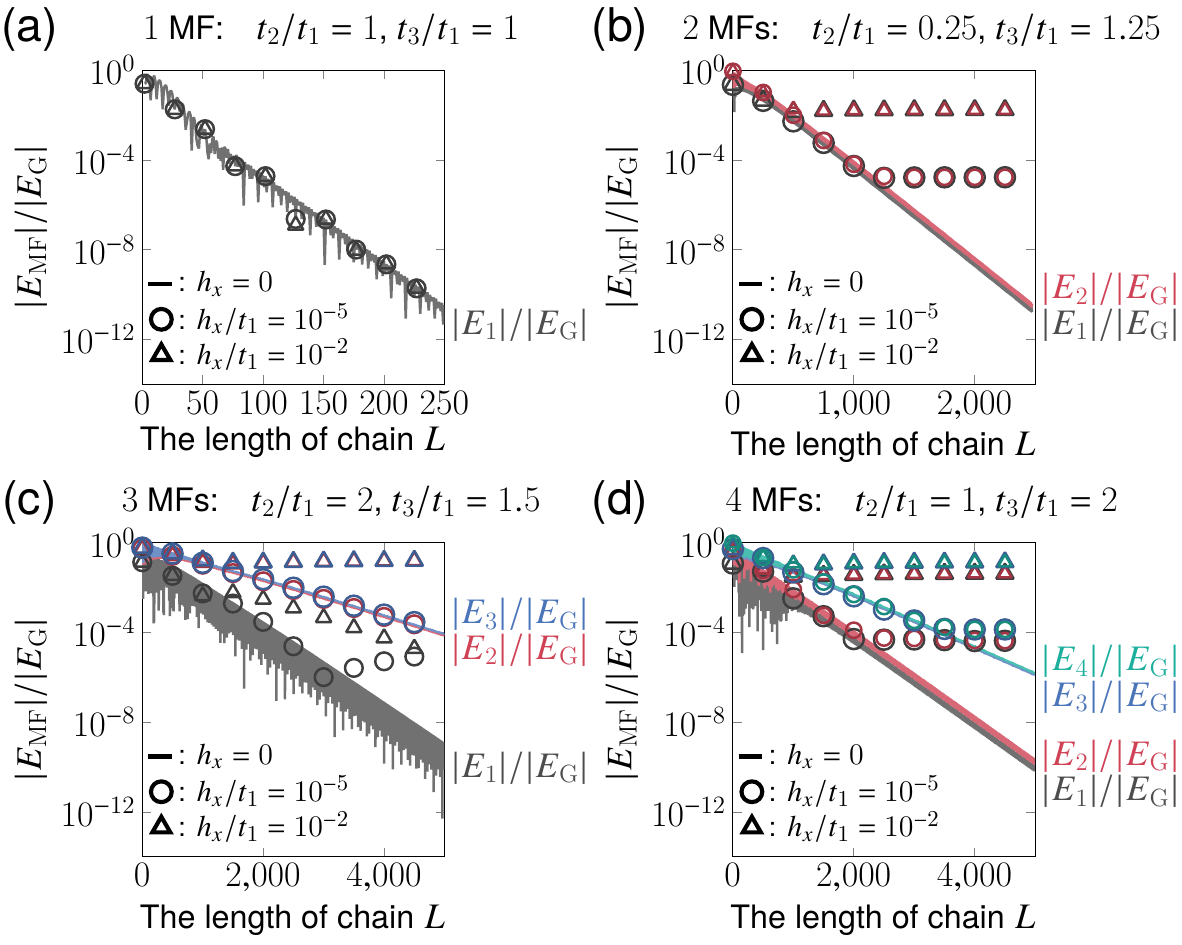}
    \caption{
    (a)--(d) Low energy spectra as a function of $L$ in the (a) 1 MF (b) 2 MF (c) 3 MF (d) 4 MF phases. For $h_x\neq 0$ ($\bigcirc$ and $\bigtriangleup$ plots), the chiral symmetry protecting Majorana fermion is broken.
    $\Delta_{\uparrow\uparrow}=0.18t_1$, $\Delta_{\downarrow\downarrow}=1.8t_1$, $\theta = \pi/4$, $h/t_1=2$, $\mu/t_1=1$.
When magnetic fields $h_x$ breaking the chiral symmetry are applied, the energy modes with the same slope gap out to the nonzero energy.
These topological phases can be distinguished by the presence of energy modes with different slopes for $h_x=0$ and the existence of an energy mode toward zero for $h_x\neq0$.
    \label{fig:energy}
    }
\end{figure}

To describe the interference of Majorana fermions at both edges, we have shown the spatial dependence of  the low energy spectra as a function of the system size in \figs{\ref{fig:energy}(a)--\ref{fig:energy}(d)}.
They have been calculated by the diagonalization of the Hamiltonian in the real space in \eq{\ref{eq:hamiltonian-real}}. 
The energy modes are labeled $|E_1|$, $|E_2|$, $|E_3|$, and $|E_4|$ in ascending order of the absolute values. 
These four values of energy correspond to the Majorana fermions, and these energy spectra are normalized by the effective energy gap $|E_\mathrm{G}|$ at each system size. 
For example, in the 1 MF (2 MF) phase, $|E_\mathrm{G}|$ is the second (third) smallest energy.

With the chiral symmetry, since multiple Majorana fermions phases can be achieved, the smallest energy modes, related to Majorana fermions, approach zero as the system size increases {[solid line plots in \figs{\ref{fig:energy}(a)--\ref{fig:energy}(d)}]}.
In the 1 MF phase, shown in \fig{\ref{fig:energy}(a)}, $|E_1|$ decreases and approaches zero with an oscillation as the system size increases.
In the 2 MF phase, shown in \fig{\ref{fig:energy}(b)}, $|E_1|$ and $|E_2|$ go to zero with degeneracy. These two energy modes are less oscillating.
The three energy modes in the 3 MF phases show different behavior compared to the 2 MF phases as shown in \fig{\ref{fig:energy}(c)}.
The energy $|E_1|$ approaches zero faster than $|E_2|$ and $|E_3|$ with the increase of the system size, and it has a large fluctuation while these degenerate $|E_2|$ and $|E_3|$ have little oscillation.
In \fig{\ref{fig:energy}(d)}, the four energy modes in the 4 MF phase are divided into two ones, and the energy ones $|E_1|$ and $|E_2|$ are fluctuating for $L<1000$.

The $\bigcirc$ and $\bigtriangleup$ plots in \fig{\ref{fig:energy}} depict the energy spectra as a function of the system size for $h_x/t_1=10^{-5},10^{-2}$. 
The magnetic field $h_x$ breaks the Majorana fermions that are protected by the chiral symmetry.
In the 1 MF phase shown in \fig{\ref{fig:energy}(a)}, the energy spectrum does not change regardless of the magnitude of the perturbation.
In the 2 MF phase, the degenerate energy modes do not go to zero but converge to a nonzero value as shown in \fig{\ref{fig:energy}(b)}.
The converged value is on the order of $|E_\mathrm{MF}|/|E_\mathrm{G}|\sim h_x/t_1$.
In the 3 MF phase, when a large perturbation is applied, the two degenerate energy modes become constant with the increase of $L$ {[\fig{\ref{fig:energy}}(c)]}.
The nonzero value is about $|E_\mathrm{MF}|/|E_\mathrm{G}|\sim 10h_x/t_1$.
Although the other energy mode changes to a gentle slope, it decreases toward zero.
\headfig{\ref{fig:energy}(d)}, in the 4 MF phase, shows that the four energy modes converge to nonzero values.
These converged values are on the order of $|E_\mathrm{MF}|/|E_\mathrm{G}|\sim h_x/t_1$ or $10h_x/t_1$.
From these results, it can be seen that it is not easy to detect the zero-energy state in the phase with an even number of Majorana fermions when magnetic fields are unintentionally applied in the $x$ direction that breaks the chiral symmetry.

\begin{figure}[!htbp]
    \centering
    \includegraphics[width=0.49\textwidth]{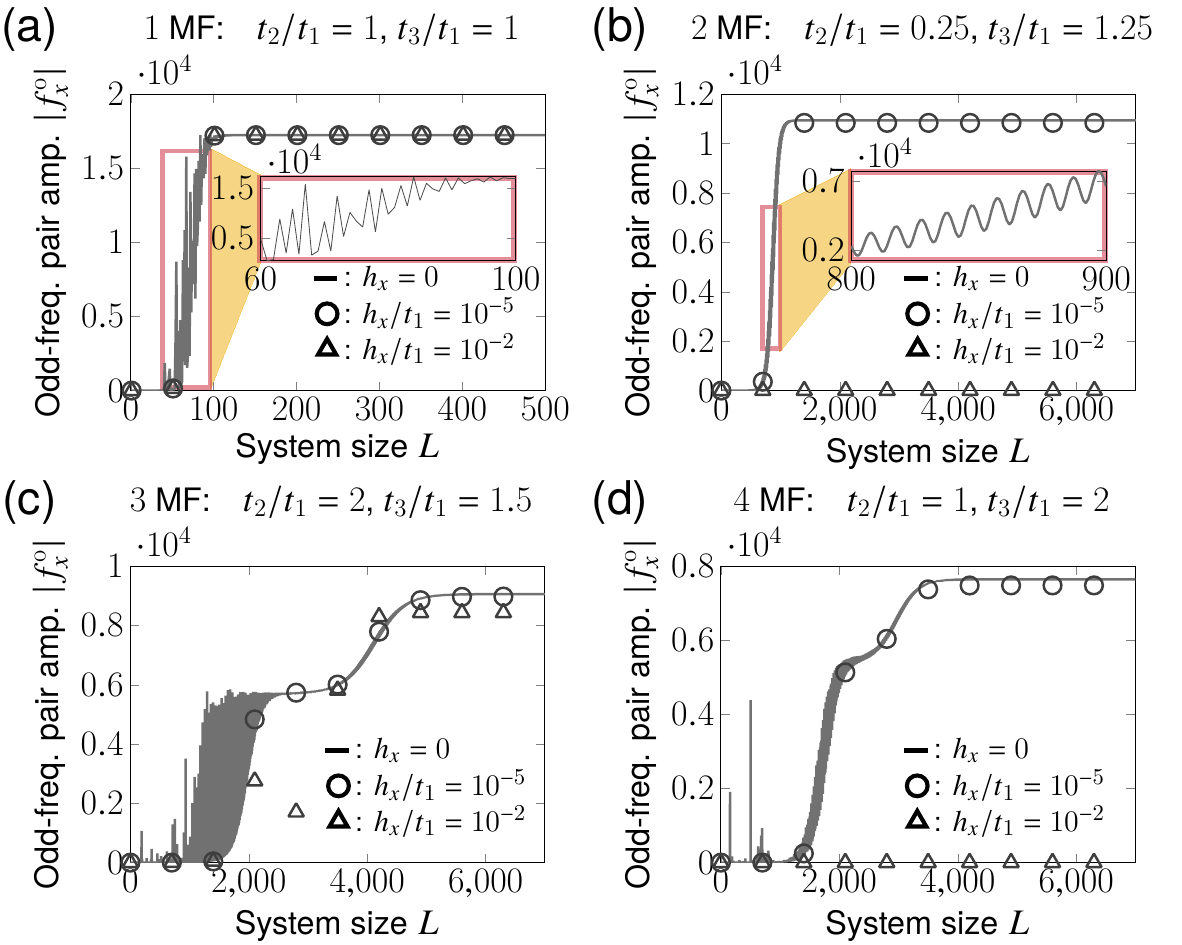}
    \caption{
    The $x$ component of the odd-frequency pair amplitude at the edge $j=1$ as a function of the system size $L$ in the (a) 1 MF (b) 2MFs (c) 3MFs (d) 4MFs phases. 
    $\theta = \pi/4$, $\Delta_{\uparrow\uparrow}=0.18t_1$, $\Delta_{\downarrow\downarrow}=1.8t_1$, $h/t_1=2$, $\mu/t_1=1$, $\omega/t_1=10^{-5}$. 
    For $h_x\neq0$ ($\bigcirc$ and $\bigtriangleup$ plots), the chiral symmetry protecting Majorana fermions is broken.
We only focus on the $x$ component of the odd-frequency pair amplitude. 
This is because the relationship between this component and the winding number is shown by the spectral bulk-boundary correspondence~[\fig{\ref{fig:sbbc}}, \headapp{\ref{sect:app_sbbc}}].
    (b) In the 2MF phase, the odd-frequency pair amplitude oscillates periodically while (c) it does randomly in the 1MF phase.
    These topological phases can be distinguished by the number of plateaus for $h_x=0$ and the existence of the odd-frequency pair amplitude for the large magnetic fields ($h_x=10^{-2}$).
For the weak magnetic fields ($h_x=10^{-4}$), these original plateau structures are maintained even though the chiral symmetry is broken.
    \label{fig:odd-freq}
    }
\end{figure}

To correlate the multiple Majorana fermions and odd-frequency Cooper pairs, we demonstrate the odd-frequency pair amplitude that is enhanced by Majorana fermions.
According to a spectral bulk-boundary correspondence, the $x$ component of the odd-frequency $\bm{\mathrm{f}}$ vector is closely related to the number of Majorana fermions in this system [\headapp{\ref{sect:app_sbbc}}]. 
For that reason, we have focused on the $x$ component of the odd-frequency $\bm{\mathrm{f}}$ vector as a function of the system size in 1, 2, 3, and 4 MF phases [\figs{\ref{fig:odd-freq}(a)--\ref{fig:odd-freq}(d)}].
These values are calculated by the Matsubara Green's function as written in \eqs{(\ref{eq:matsubara}), (\ref{eq:odd-freq}), and (\ref{eq:recurrence-left})}.

We first explain the case where the system has chiral symmetry [the solid lines in \figs{\ref{fig:odd-freq}(a)--\ref{fig:odd-freq}(d)}], similar to the energy spectra shown in \fig{\ref{fig:energy}}.
In the 1 MF phase, the odd-frequency pair amplitude increases sharply near $L=50$ with a random oscillation, and saturates for the large $L$ as shown in \fig{\ref{fig:odd-freq}(a)}.
Although the 2 MF case is similar to the 1MF one, the pair amplitude is enhanced with a periodic oscillation {[\fig{\ref{fig:odd-freq}(b)}]}.
In the 3 MF and 4 MF cases, these odd-frequency pair amplitudes have two plateaus, and they are fluctuating before the first plateaus as shown in \figs{\ref{fig:odd-freq}(c) and \ref{fig:odd-freq}(d)}.
These two plateaus are corresponding to the two energy modes shown in \figs{\ref{fig:energy}(c) and \ref{fig:energy}(d)}, and they tell us the existence of multiple Majorana phases {[We discuss the relationship between the plateau structure and the low energy spectra being related to the Majorana fermions in \headapp{\ref{sect:app_plateaus}}]}.

Additionally, we have investigated the change in the odd-frequency pair amplitude when the chiral symmetry is broken. 
The $\bigcirc$ and $\bigtriangleup$ plots in \figs{\ref{fig:odd-freq}(a)--\ref{fig:odd-freq}(d)} show how the odd-frequency pair amplitude at the edge of the system changes with increasing the system size for $h_x/t_1=10^{-5}$ and $10^{-2}$, respectively.
When the order of the perturbations $h_x$ is similar to that of the Matsubara frequency $\omega$, the single plateaus in the 1 MF and 2 MF phases and the double plateaus in the 3 MF and 4 MF phases are still kept {[the $\bigcirc$ plot in \figs{\ref{fig:odd-freq}(a)--\ref{fig:odd-freq}(d)}].
    Even if small magnetic fields that break chiral symmetry are applied, the odd-frequency pair amplitude at a finite frequency tells us the remnants of the multiple Majorana fermions. 

For the large perturbations compared to the Matsubara frequency, the single plateau in the 1 MF phase does not change {[the $\bigtriangleup$ plot in \fig{\ref{fig:odd-freq}(a)}]}.
The two plateaus in the 3 MF phase are transformed into one plateau structure {[the $\bigtriangleup$ plot in \fig{\ref{fig:odd-freq}(c)}]}.
The odd-frequency pair amplitude, in the even number of Majorana fermions phases, is suppressed, and the single and double plateaus vanish {[the $\bigtriangleup$ plot in \figs{\ref{fig:odd-freq}(b) and \ref{fig:odd-freq}(d)}]}.
It is because that the energy modes, enhancing the odd-frequency Cooper pairs, with the same localization length are mixed by the chiral symmetry breaking and the degeneracy at the almost zero energy can be solved [This can be read from \figs{\ref{fig:energy}(c) and \ref{fig:energy}(d)}].

\section{Spatial dependence of odd-frequency pair amplitude}
\label{sect:spatial}
\begin{figure*}[!htbp]
    \centering
    \includegraphics[width=\textwidth]{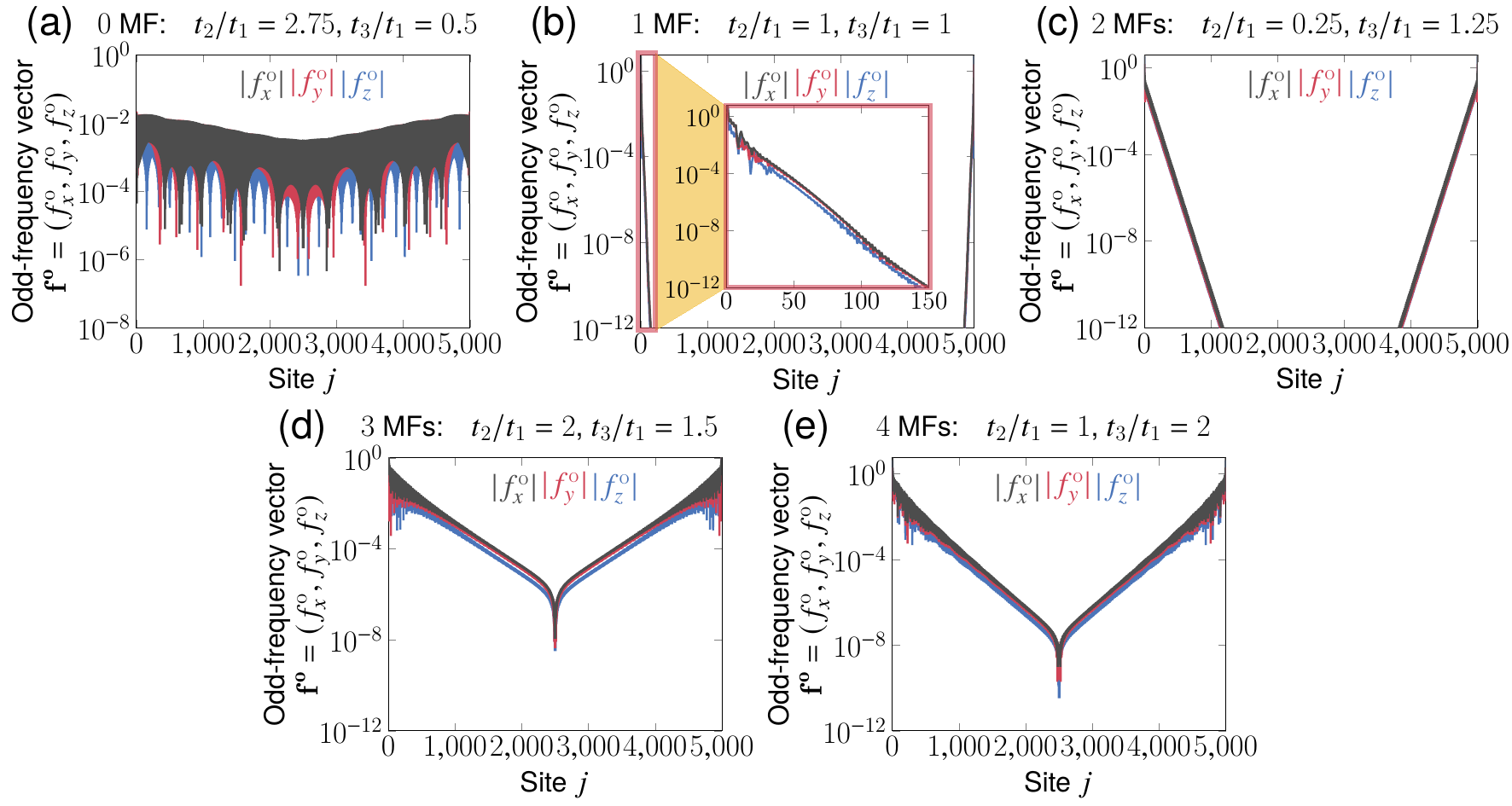}
    \caption{
        The spatial dependence of the odd-frequency $\bm{\mathrm{f}}$ vector in the (a) 0 MF (b) 1 MF (c) 2 MF (d) 3 MF (e) 4 MF phases.
    $\theta = \pi/4$, $\Delta_{\uparrow\uparrow}/t_1=0.18$, $\Delta_{\downarrow\downarrow}/t_1=1.8$, $h/t_1=2$, $\mu/t_1=1$, $L=5000$. 
    $\omega/t_1=E_G/2$: (a) $2.3\times 10^{-3}$, (b) $3.3\times 10^{-2}$, (c) $2.9\times 10^{-2}$, (d) $1.2\times 10^{-2}$, (e) $1.4\times 10^{-2}$.
        The odd-frequency pair amplitude in the topological phases [(b)--(e)] are localized at the edge of the system while that in the trivial phase [(a)] is not.
        These spatial extents depend on the gap size in the bulk shown in \fig{\ref{fig:energy-super}}.
    \label{fig:spatial-odd-freq}
    }
\end{figure*}

\begin{figure*}[!htbp]
    \centering
    \includegraphics[width=\textwidth]{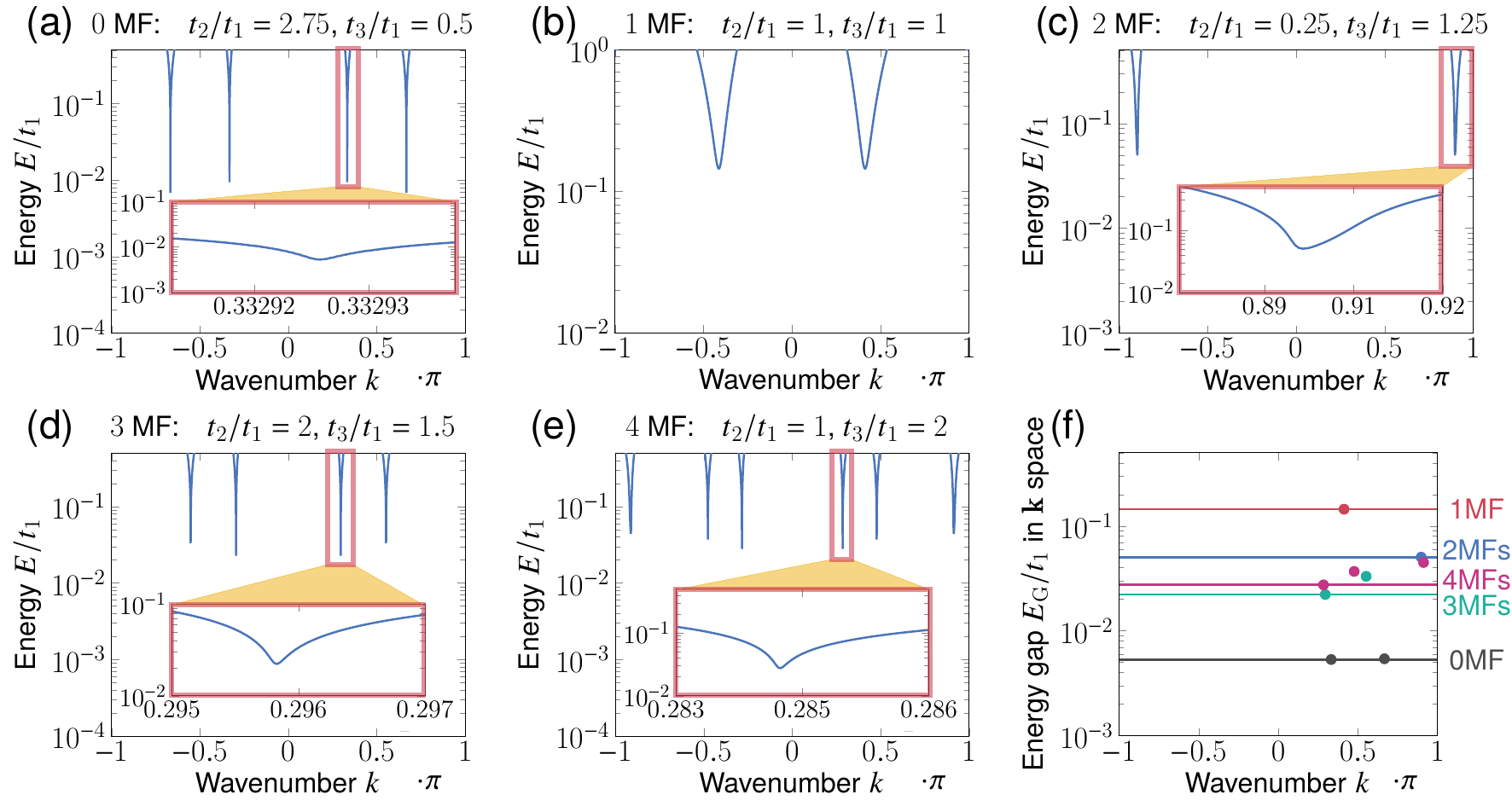}
    \caption{
        Energy bands ($E>0$) of superconducting states in $\bm{\mathrm{k}}$ space. (a) 0 MF, (b) 1 MF, (c) 2 MF, (d) 3 MF, (e) 4 MF phases. (f) The comparison of the lowest energy states in each phase.
        $\theta = \pi/4$, $\Delta_{\uparrow\uparrow}/t_1=0.18$, $\Delta_{\downarrow\downarrow}/t_1=1.8$,  $h/t_1=2$, $\mu/t_1=1$.
            The odd-frequency pair amplitude shown in \fig{\ref{fig:spatial-odd-freq}} spreads spatially as the gap size in the bulk being plotted in (f) decreases .
        \label{fig:energy-super}
    }
\end{figure*}

\begin{figure*}[!htbp]
    \centering
    \includegraphics[width=0.84\textwidth]{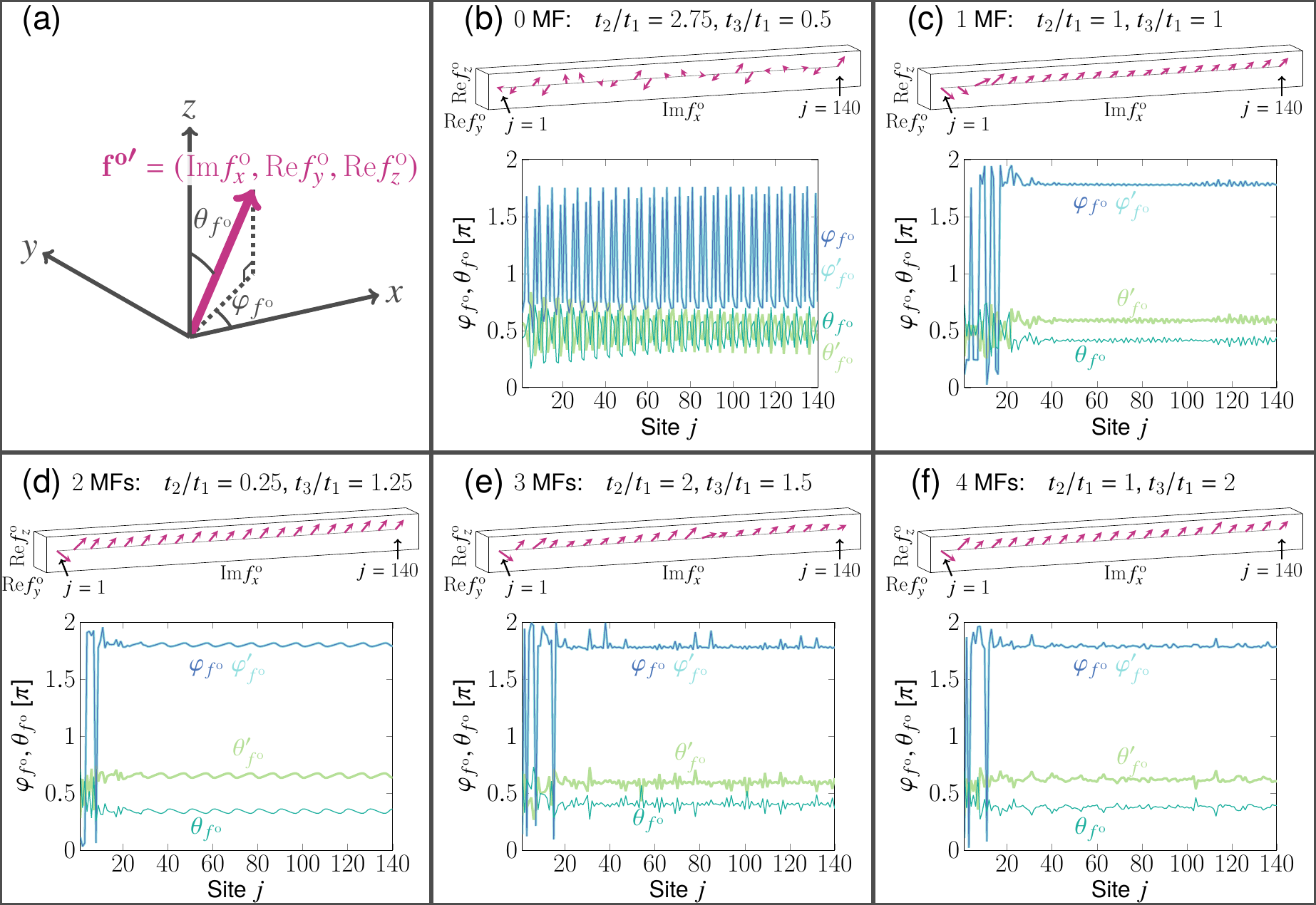}
    \caption{
        The direction of the odd-frequency $\bm{\mathrm{f}}$ vector: (a) $\fop=[\mathrm{Im}f^\mathrm{o}_x,\mathrm{Re}f^\mathrm{o}_y,\mathrm{Re}f^\mathrm{o}_z]$$=[\|\fop\| \sin\theta_{f^\mathrm{o}}\cos\varphi_{f^\mathrm{o}}, \|\fop\|\sin\theta_{f^\mathrm{o}}\sin\varphi_{f^\mathrm{o}}, \|\fop\|\cos\theta_{f^\mathrm{o}}]$ in the (b) 0 MF (c) 1 MF (d) 2 MF (e) 3 MF (f) 4 MF phases.
        The unit vector $\fop/\|\fop\|$ (upper figure) and the angle of the vector $\varphi_{f^\mathrm{o}},\theta_{f^\mathrm{o}}$ (lower figure) are plotted as a function of site $j$.
        We focus on near the edge of the system ($j\in[1,140]$).
        $\theta = \pi/4$, $\Delta_{\uparrow\uparrow}/t_1=0.18$, $\Delta_{\downarrow\downarrow}/t_1=1.8$, $h/t_1=2$, $\mu/t_1=1$, $L=5000$. 
        $\omega/t_1=E_G/2$: (a) $2.3\times 10^{-3}$, (b) $3.3\times 10^{-2}$, (c) $2.9\times 10^{-2}$, (d) $1.2\times 10^{-2}$, (e) $1.4\times 10^{-2}$.
        $\varphi'_{f^\mathrm{o}}$ and $\theta'_{f^\mathrm{o}}$: $\varphi_{f^\mathrm{o}}$ and $\theta_{f^\mathrm{o}}$ when $h_y\to -h_y$.
            These $\bm{\mathrm{f}}$ vectors are fixed in the same direction in any topological phases.
            Inverting the sign of $h_y$ reverses the sign of $\mathrm{Re}f^\mathrm{o}_z$: $\theta_{f^\mathrm{o}}'=\pi - \theta_{f^o}$.
        \label{fig:direction-odd-freq}
    }
\end{figure*}

\begin{figure*}[!htbp]
    \centering
    \includegraphics[width=0.84\textwidth]{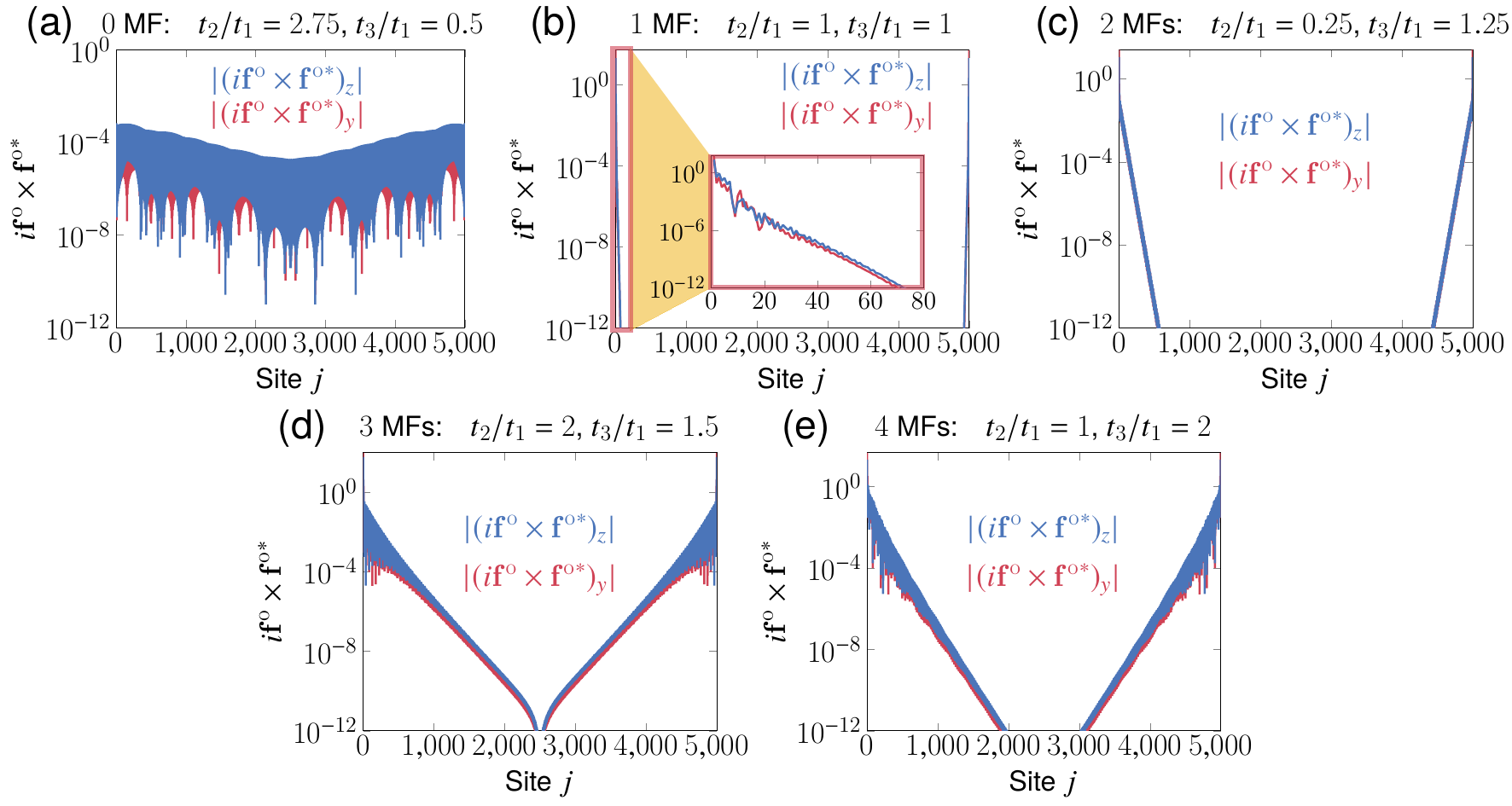}
    \caption{
        The spatial dependence of the spin states of the odd-frequency Cooper pairs $\langle\sigma^\mathrm{o}\rangle\propto i \fo\times{\fo}^*$. (a) 0 MF, (b) 1 MF, (c) 2 MF, (d) 3 MF, (e) 4 MF phases.
        $\theta = \pi/4$, $h/t_1=2$, $\mu/t_1=1$.
        $\omega/t_1=E_G/2$: (a) $2.3\times 10^{-3}$, (b) $3.3\times 10^{-2}$, (c) $2.9\times 10^{-2}$, (d) $1.2\times 10^{-2}$, (e) $1.4\times 10^{-2}$.
            The magnitude of $i\bm{\mathrm{f}}^\mathrm{o}\times{\bm{\mathrm{f}}^\mathrm{o}}^\ast$ is on the order of the square of the odd-frequency pair amplitude.
            The spin of the pair potential pointing in the $z$ direction is tilted by the magnetic fields in the $y$-$z$ plane.
            Therefore, the $x$ component becomes zero.
    \label{fig:spin-odd-freq}
    }
\end{figure*}

The next important thing to understand the relationship between the multiple Majorana fermions and odd-frequency Cooper pairs is the spin texture of odd-frequency pair amplitude. 
To evaluate the spin texture, we define the $\bm{\mathrm{f}}$ vector of the odd-frequency pairing as \eq{\ref{eq:odd-even-freq-vector}}, as well as the $\bm{\mathrm{d}}$ vector of the spin-triplet superconducting pair potential. 
In this section, we analyze the spatial dependence of the odd-frequency $\bm{\mathrm{f}}$ vector and the spin texture of it.

\headfig{\ref{fig:spatial-odd-freq}} shows the spatial dependence of the odd-frequency $\bm{\mathrm{f}}$ vector. It is calculated by the recursive Green's function's formulation {[\eqs(\ref{eq:recurrence-right}), (\ref{eq:recurrence-left}), and (\ref{eq:connect})]}.
In the 0 MF phase, the components of the odd-frequency $\bm{\mathrm{f}}$ vector are small, and spreads throughout the system with oscillations as shown in \fig{\ref{fig:spatial-odd-freq}(a)}.
In the 1 and 2 MF phases, the odd-frequency pair amplitudes are strongly localized at the edge of the system [\figs{\ref{fig:spatial-odd-freq}(b) and \ref{fig:spatial-odd-freq}(c)}].
On the other hand, in the 3 and 4 MF phases, the pair amplitudes have large values with oscillations at the edge, and it spreads toward the center compared to in the 1 and 2 MF phases [\figs{\ref{fig:spatial-odd-freq}(d) and \ref{fig:spatial-odd-freq}(e)}].

These behaviors of the odd-frequency pair amplitudes can be explained by the size of the superconducting gap in $\bm{\mathrm{k}}$ space.
This is because the spread of the odd-frequency pair amplitude in a superconducting state is about the coherence length $\xi\propto1/\Delta$ where $\Delta$ indicates the superconducting gap.
\headfigs{\ref{fig:energy-super}(a)--\ref{fig:energy-super}(e)} depict the electron part ($E>0$) of energy dispersions in 1--4 MF phases.
These energy bands are calculated by the diagonalization of the Hamiltonian [\eq{\ref{eq:hamiltonian-k-two}}] under a periodic boundary condition.
The smallest gaps in the 0, 1, 3, and 4 MF phases are located near $k=1$ while that in the 2 MF phase be near $k=\pi$.
We also show a comparison of the superconducting gap sizes in \fig{\ref{fig:energy-super}(f)}.
Comparing \figs{\ref{fig:spatial-odd-freq}(a)--(e)} and \fig{\ref{fig:energy-super}}, we can find that the magnitude of the spread of the odd-frequency pair amplitude is in ascending order of the minimum gap.

To see anisotropies of the odd-frequency pair amplitude, we illustrate the direction of the odd-frequency $\bm{\mathrm{f}}$ vectors $\fop = [\mathrm{Im}f_x^\mathrm{o},\mathrm{Re}f_y^\mathrm{o},\mathrm{Re}f_z^\mathrm{o}]$ in \fig{\ref{fig:direction-odd-freq}(a)}. The spatial dependence of the $\bm{\mathrm{f}}$ vectors, normalized by themselves, is shown on upper panels in \figs{\ref{fig:direction-odd-freq}(b)--\ref{fig:direction-odd-freq}(f)}.
Also, we depict the angle of the vectors $\theta_{f^\mathrm{o}}, \{\tan\theta_{f^\mathrm{o}} = \sqrt{(\mathrm{Im}f_x^\mathrm{o})^2+(\mathrm{Re}f_y^\mathrm{o})^2}/\mathrm{Re}f_z^\mathrm{o}, \theta_{f^\mathrm{o}}\in [0,\pi]\}$ and $\varphi_{f^\mathrm{o}}, \{\tan\varphi_{f^\mathrm{o}}=\mathrm{Re}f_y^\mathrm{o}/\mathrm{Im}f_x^\mathrm{o},\varphi_{f^\mathrm{o}} \in[0,2\pi]\}$ on lower panels in \figs{\ref{fig:direction-odd-freq}(b)--\ref{fig:direction-odd-freq}(f)} to capture their direction clearly.
Particularly, we focus on near the edge of the system $j\in[1,140]$ to correlate the vectors to the edge states.
In the trivial (0 MF) phase, the sign of the y and z components changes frequently [\fig{\ref{fig:direction-odd-freq}}(b)].
On the other hand, in the topological (1--4 MF) phases, the sign of the $x$ component changes near the edge but $\varphi_{f^\mathrm{o}}$ and $\theta_{f^\mathrm{o}}$ become constant with small oscillations slightly away from the edge {[\figs{\ref{fig:direction-odd-freq}(c)--\ref{fig:direction-odd-freq}(f)}]}.
Moreover, the odd-frequency $\bm{\mathrm{f}}$ vector, in each topological phase, points in almost the same direction [This behavior is not seen in the even-frequency $\bm{\mathrm{f}}$ vectors as shown in \headapp{\ref{sect:app_fvector}}].
The fixed vectors point slightly positive from the $x$ axis to the $y$--$z$ plane.

To investigate the effect of the magnetic fields  on the orientation of the odd-frequency $\bm{\mathrm{f}}$ vector, we change the sign of $h_y$.
Here, the sign of $h_y$ does not depend on the winding number [\headapp{\ref{sect:app_winding}}].
The angles $\varphi_{f^\mathrm{o}}'$ and $\theta_{f^\mathrm{o}}'$ represent $\varphi_{f^\mathrm{o}}$ and $\theta_{f^\mathrm{o}}$ when the sign of $h_y$ is inverted, respectively.
As can be seen from \figs{\ref{fig:direction-odd-freq}(b)--\ref{fig:direction-odd-freq}(f)}, $\theta_{f^\mathrm{o}}$ becomes $\pi-\theta_{f^\mathrm{o}}$ for the sign inversion of $h_y$ while $\varphi_{f^\mathrm{o}}$ does not change.
In other words, the sign inversion of $h_y$ inverts the sign of $f_z^{\mathrm{o}}$.

It can be understood from the analytical calculation in the bulk that the $y$ component of the magnetic fields affects the $z$ component of the even-frequency $\bm{\mathrm{f}}$ vector.
We calculate the even-frequency spin-triplet $\bm{\mathrm{f}}$ vector in the bulk \footnote{odd-frequency spin-triplet pair amplitude does not exist in the bulk} as
\begin{widetext}
\begin{align}
    \label{eq:even-freq-k-space}
    f_x^{\mathrm{e}}=\frac{i\sin k}{A_\mathrm{denom} }\bigl[
        &(\Delta_{\uparrow\uparrow}+\Delta_{\downarrow\downarrow}) h_z^2
        + 2(\Delta_{\uparrow\uparrow}-\Delta_{\downarrow\downarrow})\varepsilon h_z 
        + (\Delta_{\uparrow\uparrow}+\Delta_{\downarrow\downarrow})h_y^2 \nonumber\\
        &+ (\Delta_{\uparrow\uparrow}+\Delta_{\downarrow\downarrow})\omega^2
        + 4\Delta_{\uparrow\uparrow}\Delta_{\downarrow\downarrow}(\Delta_{\uparrow\uparrow}+\Delta_{\downarrow\downarrow})\sin^{2} k
        + (\Delta_{\uparrow\uparrow}+\Delta_{\downarrow\downarrow})\varepsilon^2
    \bigr]\nonumber\\
    f_y^{\mathrm{e}} = \frac{\sin k}{A_\mathrm{denom} }\bigl[
        & (\Delta_{\downarrow\downarrow}-\Delta_{\uparrow\uparrow})h_z^2
        -2(\Delta_{\uparrow\uparrow}+\Delta_{\downarrow\downarrow})\varepsilon h_z
        +(\Delta_{\uparrow\uparrow}-\Delta_{\downarrow\downarrow})h_y^2\nonumber\\
        &+(\Delta_{\downarrow\downarrow}-\Delta_{\uparrow\uparrow})\omega^2
        +4\Delta_{\uparrow\uparrow}\Delta_{\downarrow\downarrow}(\Delta_{\uparrow\uparrow}-\Delta_{\downarrow\downarrow})\sin^{2} k
        +(\Delta_{\downarrow\downarrow}-\Delta_{\uparrow\uparrow})\varepsilon^2
    \bigr]\nonumber\\
    f_z^{\mathrm{e}} = \frac{\sin {k} }{A_\mathrm{denom} }
    \bigl[
        &2(\Delta_{\uparrow\uparrow}-\Delta_{\downarrow\downarrow})h_y h_z 
        +2(\Delta_{\uparrow\uparrow}+\Delta_{\downarrow\downarrow})\varepsilon h_y
    \bigr]
\end{align}
with
\begin{align*}
    A_\mathrm{denom} = & h_z^4
    + 2\bigl[h_y^2+\omega^2+2(\Delta_{\uparrow\uparrow}^2+\Delta_{\downarrow\downarrow}^2)\sin^2 k-\varepsilon^2\bigr]h_z^2
    +8(\Delta_{\uparrow\uparrow}^2-\Delta_{\downarrow\downarrow}^2)\varepsilon h_z\sin^2 k
    + h_y^4
    +2(\omega^2+4\Delta_{\uparrow\uparrow}\Delta_{\downarrow\downarrow}\sin^2 k-\varepsilon^2)h_y^2\nonumber\\
    &+\omega^4
    +2\bigl[
        2(\Delta_{\uparrow\uparrow}^2+\Delta_{\downarrow\downarrow}^2)\sin^2 k +\varepsilon^2
    \bigr]\omega^2 
    +16\Delta_{\uparrow\uparrow}^2\Delta_{\downarrow\downarrow}^2\sin^4 k
    +4(\Delta_{\uparrow\uparrow}^2+\Delta_{\downarrow\downarrow}^2)\varepsilon^2\sin^2 k
    +\varepsilon^4
\end{align*}
\end{widetext}
by taking the inverse [\eq{\ref{eq:matsubara}}] of the Hamiltonian [\eq{\ref{eq:hamiltonian-k-one}}] in $\bm{\mathrm{k}}$ space and using \eqs{(\ref{eq:odd-even-freq-vector}) and (\ref{eq:even-freq})}.
As you can see \eq{\ref{eq:even-freq-k-space}}, $f_z^{\mathrm{e}}$ is an odd function of $h_y$ while $f_x^{\mathrm{e}}$ and $f_y^{\mathrm{e}}$ are even functions of it. 
The sign change of ``$f_z^\mathrm{o}$'' vector, caused by the operation $h_y\to -h_y$, is inherited from the nature in the bulk. 
This is because a translational symmetry breaking generates odd-frequency spin-triplet even-party Cooper pairs from even-frequency spin-triplet odd-parity ones.

In this system, the nonunitary state is important for accessing the phase with multiple Majorana fermions.
The matter can be understood from the structure of the winding number that is mentioned in \headapp{\ref{sect:app_winding}}.
To investigate the nonunitary state of the odd-frequency Cooper pair $\langle\sigma^\mathrm{o}\rangle$, we calculate a cross vector $\langle\sigma^\mathrm{o}\rangle\propto i \fo\times{\fo}^*$ from the odd-frequency $\bm{\mathrm{f}}$ vector.
The cross vector takes a nonzero value for the nonunitary states.

The spatial dependence of the cross vectors in the 0--4 MF phases, shown in \figs{\ref{fig:spin-odd-freq}(a)--\ref{fig:spin-odd-freq}(e)}, are similar to that of the pair amplitudes in \figs{\ref{fig:spatial-odd-freq}(a)--\ref{fig:spatial-odd-freq}(e)}.
They are on the order of about the square of the pair amplitude.
In the first place, the spin state of pair potential $\langle\sigma\rangle\propto i {\bm{\mathrm{d} }}\times{\bm{\mathrm{d} }}^*$ is obtained as $i {\bm{\mathrm{d} }}\times{\bm{\mathrm{d} }}^* = [0, 0, (\Delta_{\uparrow\uparrow}^2-\Delta_{\downarrow\downarrow}^2)/2]$ where we assume $d_z=0$ and  $\Delta_{\uparrow\uparrow}$ and $\Delta_{\downarrow\downarrow}$ are real.
Since the magnetic fields, applied along $y=z$ direction, tilt the spin state of the odd-frequency Cooper pairs in the direction of $y=z$, only the $x$ component of $\langle\sigma^\mathrm{o}\rangle\propto i \fo\times{\fo}^*$ is zero.
This result can also be easily confirmed in a Kitaev chain [\headapp{\ref{sect:app_cross}}].

\section{Discussion}
\label{sect:discussion}

\begin{figure}[!htbp]
    \centering
    \includegraphics[width=0.49\textwidth]{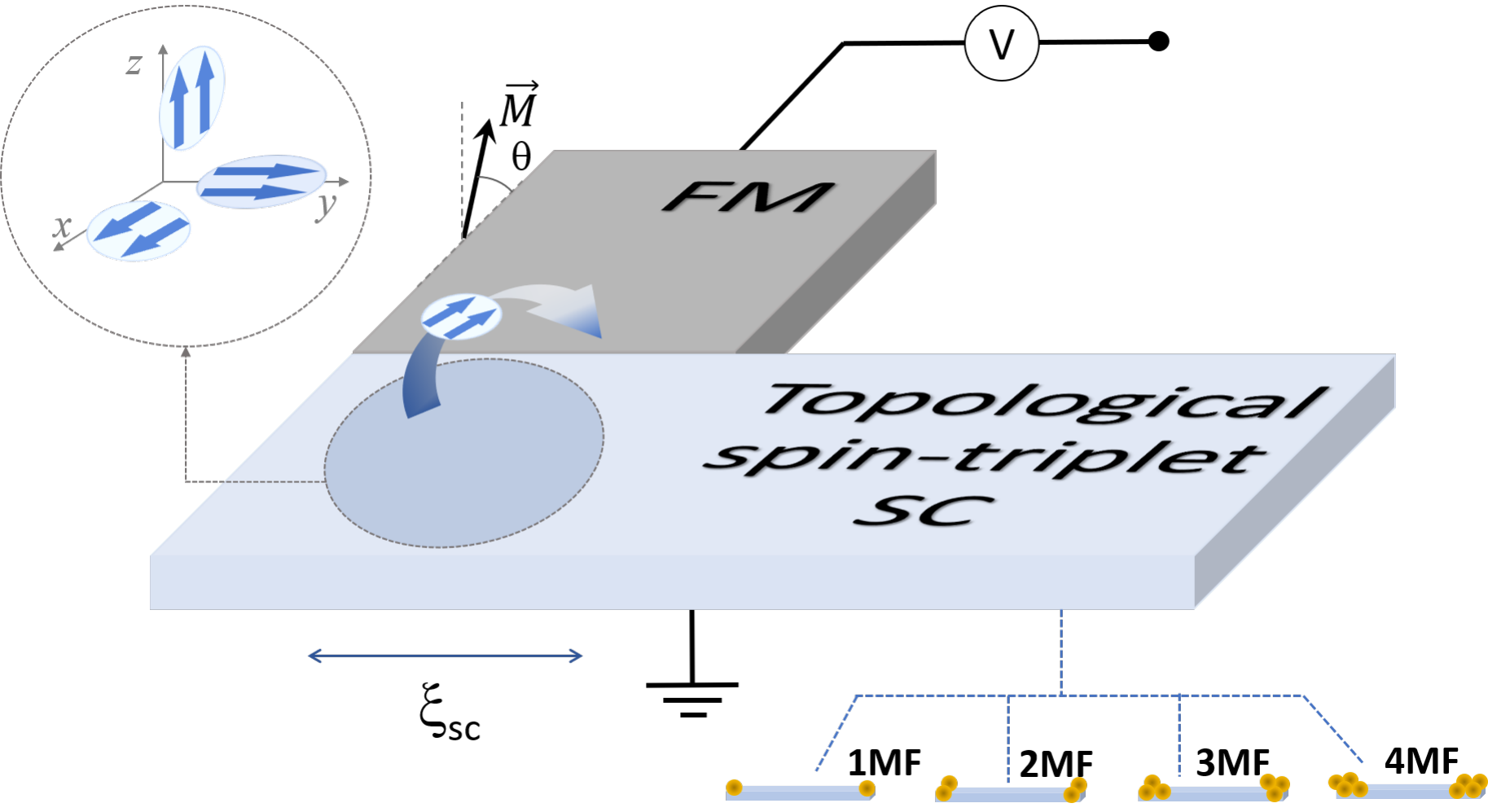}
    \caption{
    \label{fig:proposal-model}
    A schematic view for detecting spin-dependent odd-frequency Cooper pairs in a topological spin-triplet superconductor by the scanning tunneling spectroscopy. A ferromagnetic (FM) tip with impurities contacts the edge of a topological spin-triplet superconductor (SC).
    }
\end{figure}

In this section, we propose a method for detecting the spin structure of the odd-frequency Cooper pairs. 
When the spin direction of the odd-frequency spin-triplet $s$-wave Cooper pair is equal to that of the ferromagnet, it can penetrate into the ferromagnet over a long distance as compared to the case where the spin directions are different~\cite{bergeret2001long,bergeret2005odd,buzdin2005proximity,eschrig2015spin,golubov2004current,linder2015superconducting}. 
In addition, the odd-frequency spin-triplet $s$-wave Cooper pairs are robust and resonate in impurities.
This behavior has been known in the context of the anomalous proximity effect in a diffusive normal metal attached to a spin-triplet $p$-wave superconductor~\cite{tanaka2007theory,tanaka2004anomalous,tanaka2005theory,asano2006anomalous,ikegaya2016quantization,takagi2020odd}.

Taking advantage of these properties of the odd-frequency $s$-wave Cooper pairs, we propose a way to observe the electronic density of states of a topological spin-triplet superconductor by the scanning tunneling spectroscopy with a ferromagnetic chip containing impurities as shown in \fig{\ref{fig:proposal-model}}. 
Then, the local density of states is expected to have a zero-energy peak or split ones. 
By measuring the density of states at the fixed low energy at the edges of samples with different lengths and comparing the obtained data, it may be possible to show one or two plateau structures in \fig{\ref{fig:odd-freq}}.
Additionally, the structure of the odd-frequency $\bm{\mathrm{f}}$ vector with a fixed orientation, shown in \figs{\ref{fig:direction-odd-freq}(b)--\ref{fig:direction-odd-freq}(f)}, may be detected by using ferromagnetic chips with different spin directions for one sample and comparing the peak heights of the low-energy density of states.

\section{Conclusion}
\label{sect:conclusion}
In this paper, we have shown that the spatial profile of the odd-frequency spin-triplet $s$-wave Cooper pairs can become an indicator for detecting and distinguishing the topological phases with different numbers of Majorana fermions in a spin-triplet $p$-wave superconductor with magnetic fields. 
It is not easy to distinguish the topological phases with different numbers of Majorana fermions by only focusing on the local density of states at the edge of the semi-infinite system.
To solve this difficulty, we have considered the situation where the energy level of Majorana fermions deviate from zero by choosing the finite system size and applying additional magnetic fields to break the chiral symmetry.
The induced odd-frequency pair amplitude at the edge as a function of the system size has two plateaus when the two energy modes with different localization lengths exist, such as in the 3 and 4 Majorana fermions phases. 
While the weak magnetic fields, breaking the chiral symmetry, gap out the zero energy states in the even number of Majorana fermions phase, the odd-frequency pair amplitude at the edge as a function of the system size keeps its original shape. 
This result means that the odd-frequency pair amplitude tells us the fingerprints of the existence of the multiple Majorana fermions.
The strong magnetic fields that break the chiral symmetry suppress the odd-frequency pair amplitude in only the even number of Majorana fermion phases. 
The suppression allows us to distinguish between the even and odd numbers of Majorana fermions.

The structure of the winding number tells us the importance of the nonunitary superconducting states to access the multiple Majorana phases.
Therefore, we have defined the odd-frequency $\bm{\mathrm{f}}$ vector on the analogy of the $\bm{\mathrm{d}}$ vector that characterizes the spin-triplet pair potential.
Using the odd-frequency $\bm{\mathrm{f}}$ vector, we have obtained the following three results.
The odd-frequency $\bm{\mathrm{f}}$ vectors near the edge are oriented in the same direction in each topological phase, unlike the even-frequency ones. 
By inverting the sign of the magnetic fields with keeping the winding number, we have found that the magnetic response of the odd-frequency $\bm{\mathrm{f}}$ vector inherits the properties of the even-frequency one in the bulk. 
The nonunitary spin states of the odd-frequency Cooper pairs, defined by the cross vector $i\bm{\mathrm{f}}\times\bm{\mathrm{f}}^*$, tend to point in the direction of the applied magnetic fields. 
Our results mean that the presence of the nonunitary spin state of odd-frequency Cooper pairs can feature the existence of a topological phase with multiple Majorana fermions.

\section*{Acknowledgments}
We thank S. Ikegaya for fruitful discussions.
This work was supported by Grants-in-Aid from JSPS for Scientific Research (A) (KAKENHI Grant No. JP20H00131), Scientific Research (B) (KAKENHI Grants No. JP18H01176 and No. JP20H01857), Japan-RFBR Bilateral Joint Research Projects/Seminars No. 19-52-50026, and the JSPS Core-to-Core program ``Oxide Superspin'' international network.

\appendix

\appsection{Structure of the winding number}
\label{sect:app_winding}
The winding number, defined in \eq{\ref{eq:winding}}, characterizes the topology of the $p$-wave superconductor system shown in \fig{\ref{fig:system-phase-energy-normal}(a)}.
In this Appendix, we investigate the structure of the winding number in the system.

The structure of the winding number is expressed by the off-diagonal component of the off-diagonalized Hamiltonian:
\begin{align}
    U^\dagger H(k) U = 
    \begin{bmatrix}
        O & \tilde{H}\\
        \tilde{H}^\dagger & O
    \end{bmatrix},
\end{align}
where $U$ is the $2\times 2$ matrix diagonalizing the chiral operator $\Gamma=\tau_x\sigma_z$, and $H(k)$ is obtained in \eq{\ref{eq:hamiltonian-k-two}}.
When $k$ change from $-\pi$ to $\pi$, the number of times of the determinant 
\begin{align}
    \label{eq:winding-two}
    \det \tilde{H} &= \varepsilon^2(k) -h_y^2 -h_z^2 + 4\Delta_{\uparrow\uparrow}\Delta_{\downarrow\downarrow}\sin^2 k\nonumber\\
                   &\quad+ 2i \sin k [(\Delta_{\downarrow\downarrow}-\Delta_{\uparrow\uparrow})\varepsilon(k)-(\Delta_{\uparrow\uparrow}+\Delta_{\downarrow\downarrow})h_z]
\end{align}
going around the origin on the complex plane corresponds to the winding number.
Two matters can be seen from \eq{\ref{eq:winding-two}} as follows.
\begin{itemize}
    \item The winding number does not depend on the sign of $h_y$. 
    \item The nonunitary pairing ($\Delta_{\uparrow\uparrow}\neq\Delta_{\downarrow\downarrow}$) is important to reach the multiple Majorana fermion phases. [The term $2i \sin k (\Delta_{\uparrow\uparrow}-\Delta_{\downarrow\downarrow})\varepsilon(k)$ changes the sign of the imaginary part of $\det\tilde{H}$ more.]
\end{itemize}

In \sect{\ref{sect:spatial}}, we investigate the magnetic field response of the odd-frequency $\bm{\mathrm{f}}$ vector by changing the sign of $h_y$ so as not to change the phase.
Additionally, we calculate the cross vector $i \fo\times{\fo}^*$ to investigate the nonunitary state of the odd-frequency Cooper pairs. 

\appsection{Spectral bulk-boundary correspondence}
\label{sect:app_sbbc}

\begin{figure}[!htbp]
    \centering
    \includegraphics[width=0.49\textwidth]{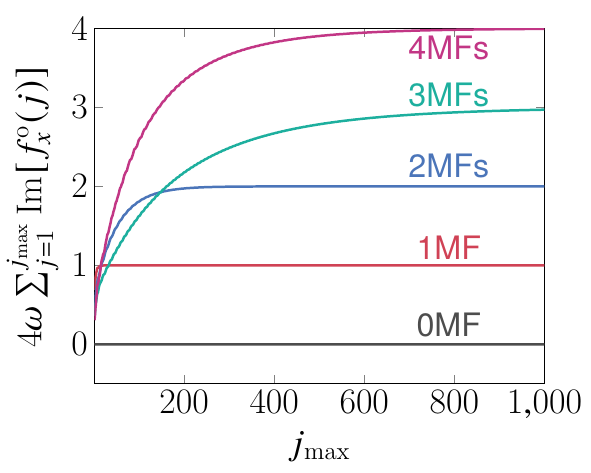}
    \caption{
        Sum of the odd-frequency $\bm{\mathrm{f}}$ vector $\mathrm{Im} f_x$ from the edge $j=1$ to $j_\mathrm{max}$ as a function of $j_\mathrm{max}$.
        This value represents $F_\mathrm{edge}$ [\eqs{(\ref{eq:sbbc-two}) and (\ref{eq:fedge})}] in the spectral bulk-boundary correspondence [\eq{\ref{eq:sbbc-one}}] in a semi-infinite $p$-wave superconductor with magnetic fields at a infinitesimal frequency.  
        $\theta = \pi/4$, $\Delta_{\uparrow\uparrow}/t_1=0.18$, $\Delta_{\downarrow\downarrow}/t_1=1.8$, $h/t_1=2$, $\mu/t_1=1$, $\omega/t_1=10^{-5}$. 0 MF: $t_2/t_1=2.75$, $t_3/t_1=0.5$, 1 MF: $t_2/t_1=1$, $t_3/t_1=1$, 2 MFs: $t_2/t_1=0.25$, $t_3/t_1=1.25$, 3 MFs: $t_2/t_1=2$, $t_3/t_1=1.5$, 4 MFs: $t_2/t_1=1$, $t_3/t_1=2$. 
            For the large $j_\mathrm{max}$ and the low frequency $\omega$, 
            $4\omega\sum_{j=1}^{j_\mathrm{max}} \mathrm{Im}[f_x^\mathrm{o}(j)]$ equals the winding number.
        \label{fig:sbbc}
    }
\end{figure}

Spectral bulk-boundary correspondence is a relationship that connects a winding number being extended to a finite frequency and the odd-frequency pair amplitudes.
The relationship is defined in a semi-infinite system with chiral symmetry, and is written as 
\begin{align}
    \label{eq:sbbc-one}
    &i\omega F_\mathrm{edge}^\mathrm{odd} (i\omega) = w_\mathrm{bulk},\\
    \label{eq:sbbc-two}
    &\quad\begin{cases}
        w_\mathrm{bulk}=\frac{i}{2}\mathrm{Tr}_k [\Gamma G(\omega)\partial_k G^{-1}(i\omega)]\\
        F_\mathrm{edge}^\mathrm{odd}(i\omega)=\mathrm{Tr}_j[\Gamma G(i\omega)],
    \end{cases}
\end{align}
with a Matsubara Green's function $G(i\omega)$  and a chiral operator $\Gamma$. For an infinitesimal Matsubara frequency, $w_\mathrm{bulk}$ becomes equal to the winding number.

In the $p$-wave superconductor with magnetic fields shown in \fig{\ref{fig:system-phase-energy-normal}(a)}, the second equation in \eq{\ref{eq:sbbc-two}} can be transformed as below:
\begin{align}
    F_\mathrm{edge}^\mathrm{odd} &= \sum_j[F_{j,j,\uparrow,\uparrow}-F_{j,j,\downarrow,\downarrow}+\tilde{F}_{j,j,\uparrow,\uparrow}-\tilde{F}_{j,j,\downarrow,\downarrow}]\nonumber\\
    \label{eq:fedge}
                                 &= 4\omega \sum_j \mathrm{Im} f_x^\mathrm{o} (j,z)
\end{align}
where we use the chiral operator $\Gamma=\tau_x\sigma_z$ anticommuting with the Hamiltonian $H(k)$ in \eq{\ref{eq:hamiltonian-k-two}}, the relation $\tilde{F}=-F^*$, and \eqs{(\ref{eq:matsubara})--(\ref{eq:odd-freq})}.
The sum of the odd-frequency vector $\mathrm{Im} f_x$ from the edge $j=1$ to $j_\mathrm{max}$ [$F_\mathrm{edge}$ in \eq{\ref{eq:fedge}}] is plotted as a function of $j_\mathrm{max}$ in \fig{\ref{fig:sbbc}}. For a sufficiently large $j_\mathrm{max}$, $w_\mathrm{bulk}$ matches the winding number (the number of Majorana fermions).
In \sect{\ref{sect:odd-freq}}, we focus on $f_x^\mathrm{o}$ among the components of the odd-frequency $\bm{\mathrm{f}}$ vector.
This is because this component is most closely related to the winding number that is extended to a finite frequency.

\appsection{Relationship between the low-energy spectra being related to Majorana fermions and the plateaus of odd-frequency pair amplitude}
\label{sect:app_plateaus}

\begin{figure}[!htbp]
    \centering
    \includegraphics[width=0.45\textwidth]{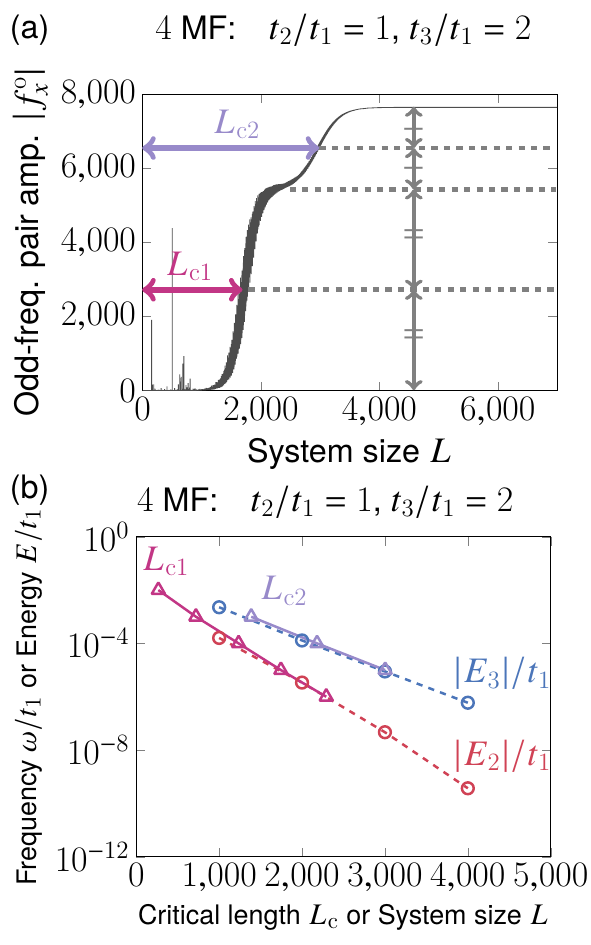}
    \caption{
        \label{fig:plateau}
    (a) Definition of the localization length of the odd-frequency pair amplitude.
    The $x$ component of the odd-frequency pair amplitude at the edge in the 4 MF phase {[\fig{\ref{fig:odd-freq} (d)}]} is replotted.
The localization lengths $L_{\mathrm{c}1}$ and $L_{\mathrm{c}2}$ of the odd-frequency pair amplitude are defined as the system sizes taking half the heights of the steps.
    Matsubara frequency: $\omega/t_1=10^{-5}$. 
    (b) Comparison between the localization length of the odd-frequency pair amplitude {[the $\bigtriangleup$ plots with the solid lines]} and the system size dependence of the low energy modes being associated with the Majorana fermions {[the $\bigcirc$ plots with the dashed plots]}.   
    The energy modes $|E_2|$ and $|E_3|$ are selected from the four modes in the 4 MF phase shown in \fig{\ref{fig:odd-freq} (d)}.
    The slopes of $|E_2|/t_1$ and $|E_3|/t_1$ correspond to that of $L_{\mathrm{c}1}$ and $L_{\mathrm{c}2}$, respectively.
    $\theta = \pi/4$, $\Delta_{\uparrow\uparrow}=0.18t_1$, $\Delta_{\downarrow\downarrow}=1.8t_1$, $h/t_1=2$, and $\mu/t_1=1$.
    }
\end{figure}

In this section, we discuss the relationship between the low energy spectra being related to the Majorana fermions  {[\fig{\ref{fig:energy}}, \sect{\ref{sect:odd-freq}}]} and the plateaus of the odd-frequency pair amplitude {[\fig{\ref{fig:odd-freq}}, \sect{\ref{sect:odd-freq}}]}.
To define the localization lengths of the odd-frequency pair amplitude, we replot \fig{\ref{fig:odd-freq} (d)} in \fig{\ref{fig:plateau} (a)}. 
Here, we focus on the odd-frequency pair amplitude in the 4 MF phase with less oscillation than that in the 3 MF phase.
The localization lengths $L_{\mathrm{c}1}$ and $L_{\mathrm{c}2}$ of the odd-frequency pair amplitude are defined as the system sizes taking half the heights of the steps (plateaus), as shown in \fig{\ref{fig:plateau} (a)}.
These lengths depend on the Matsubara frequency.

    \headfig{\ref{fig:plateau} (b)} shows the system size dependence of the low energy modes being related to the Majorana fermions {[the $\bigcirc$ plots with the dashed lines]} and the localization lengths of the odd-frequency pair amplitude as a function of the Matsubara frequency {[the $\bigtriangleup$ plots with the solid lines]}.
To compare their behaviors, we set the localization length on the $x$-axis and the frequency on the $y$-axis.
In the process of calculating the localization length, we remove its oscillation by cutting off the high-frequency component by the fast Fourier transform.
We select energy modes $|E_2|$ and $|E_3|$ from the four modes shown in \fig{\ref{fig:energy} (d)} because these two modes have less oscillation and different tilts.
Unlike \fig{\ref{fig:energy}}, the energy modes in \fig{\ref{fig:plateau} (b)} are normalized by $t_1$. 

As shown in \fig{\ref{fig:plateau} (b)}, the localization lengths $L_{\mathrm{c}1}$ and $L_{\mathrm{c}2}$ of the odd-frequency pair amplitude increase as the Matsubara frequency $\omega$ decreases. 
The slopes of these localization lengths $L_{\mathrm{c}1}$ and $L_{\mathrm{c}2}$ almost correspond to that of the energy spectra $|E_2|$ and $|E_3|$, respectively.
This means that the plateau structure of the odd-frequency pair amplitude appears when the energy becomes smaller than the fixed Matsubara frequency.
For example, the double plateau structure emerges in the plot up to $L=4000$ {[\fig{\ref{fig:plateau} (a)}]} because the energy modes $|E_2|/t_1\sim10^{-10}$ and $|E_3|/t_1\sim10^{-7}$ at $L=4000$ in \fig{\ref{fig:plateau} (b)} become smaller than the fixed Matsubara frequency $\omega/t_1=10^{-5}$ in \fig{\ref{fig:plateau} (a)}.
This result supports the argument for distinguishing the multiple Majorana phases by the odd-frequency pair amplitude in \sect{\ref{sect:odd-freq}}.

\appsection{Spatial dependence of even-frequency pair amplitude}
\label{sect:app_fvector}

\begin{figure*}[!htbp]
    \centering
    \includegraphics[width=0.84\textwidth]{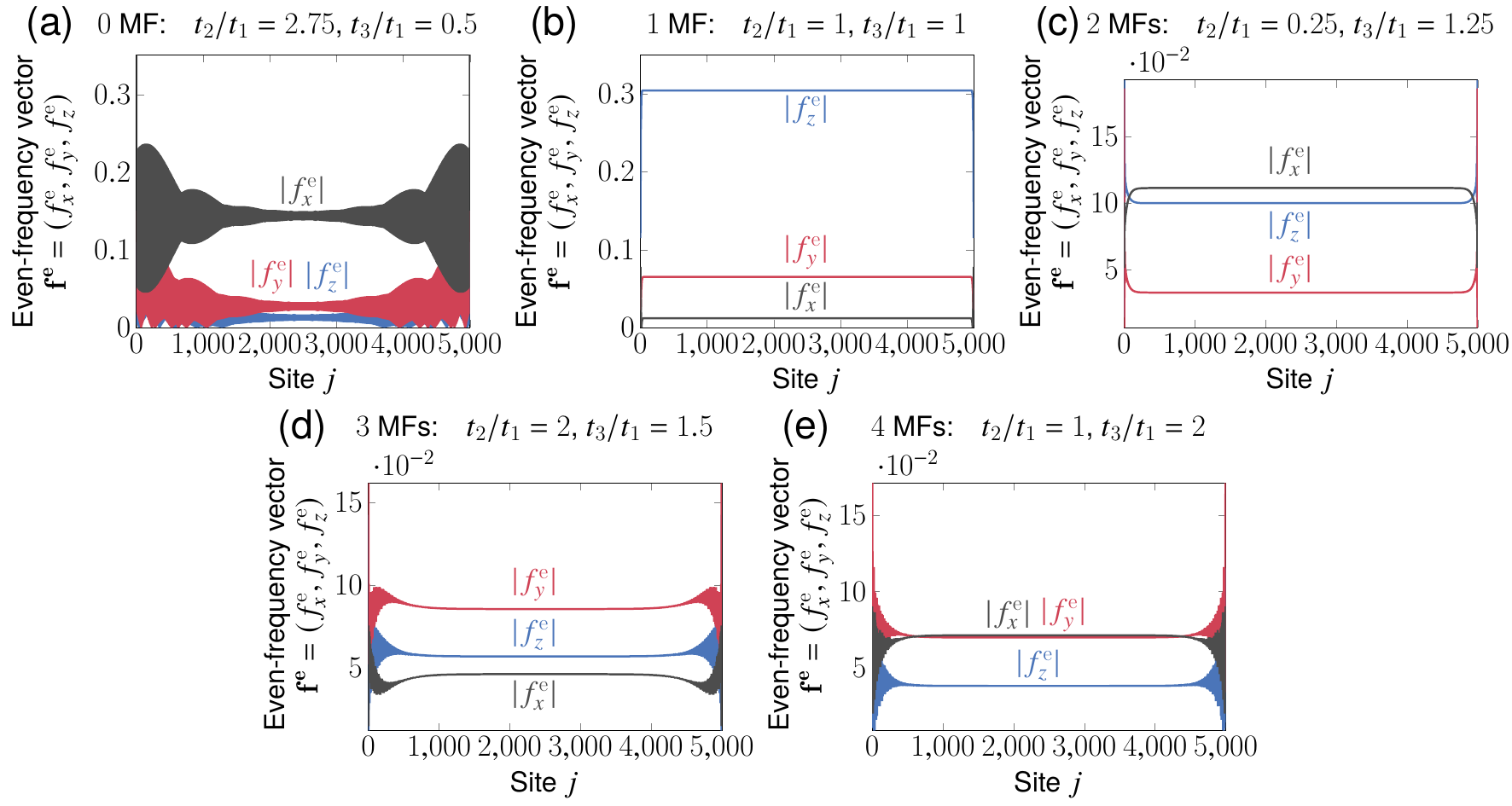}
    \caption{
        The spatial dependence of the even-frequency $\bm{\mathrm{f}}$ vector in the (a) 0 MF (b) 1 MF (c) 2 MF (d) 3 MF (e) 4 MF phases.
    $\theta = \pi/4$, $\Delta_{\uparrow\uparrow}/t_1=0.18$, $\Delta_{\downarrow\downarrow}/t_1=1.8$, $h/t_1=2$, $\mu/t_1=1$, $L=5000$. 
    $\omega/t_1=E_G/2$: (a) $2.3\times 10^{-3}$, (b) $3.3\times 10^{-2}$, (c) $2.9\times 10^{-2}$, (d) $1.2\times 10^{-2}$, (e) $1.4\times 10^{-2}$.
    Unlike the odd-frequency pair amplitude, the even-frequency pair amplitude exists in the center of the system.
    \label{fig:spatial-even-freq}
    }
\end{figure*}

\begin{figure*}[!htbp]
    \centering
    \includegraphics[width=0.84\textwidth]{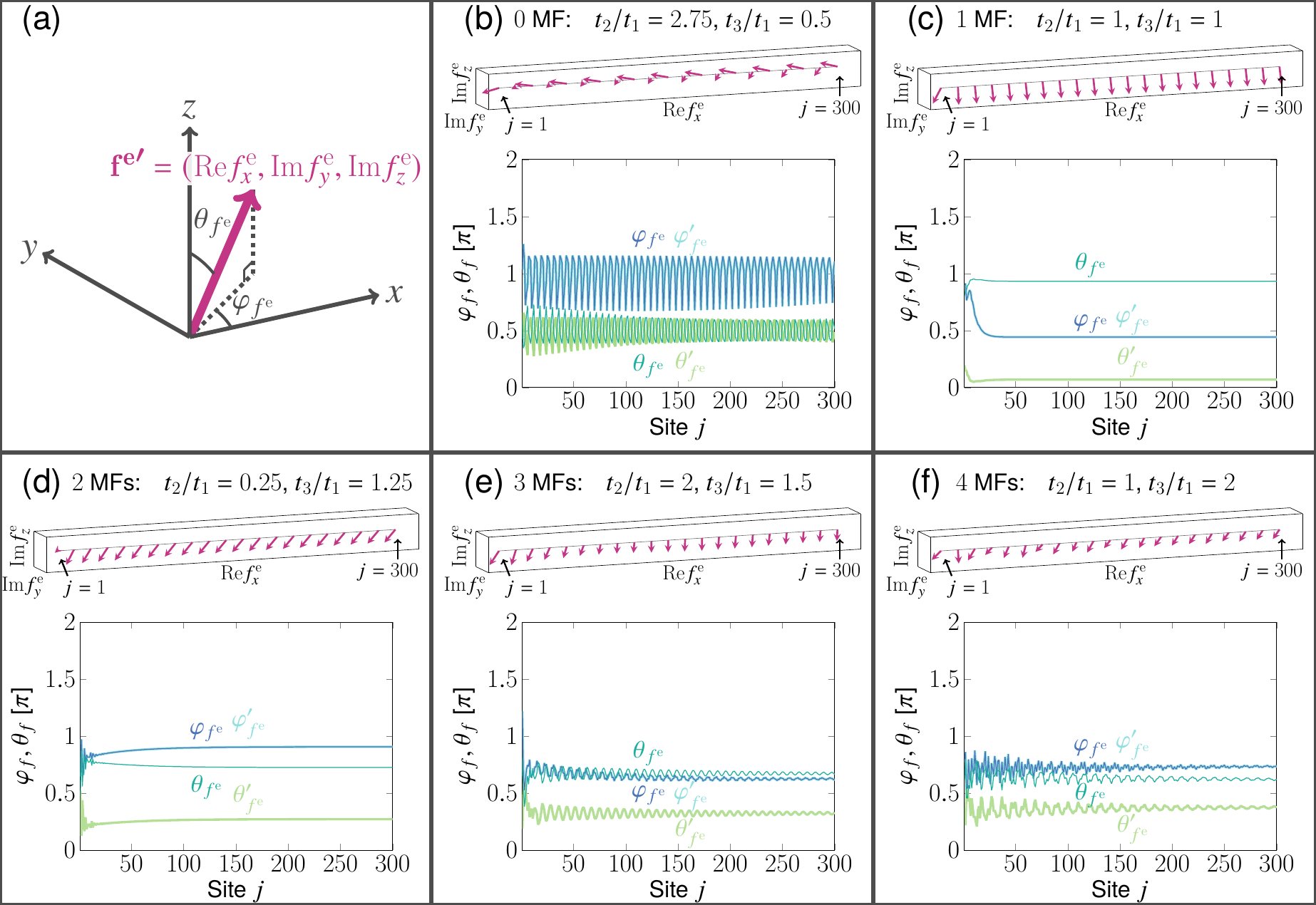}
    \caption{
        The direction of the even-frequency $\bm{\mathrm{f}}$ vector: (a) $\fep=$$[\mathrm{Re}f^\mathrm{e}_x,\mathrm{Im}f^\mathrm{e}_y,\mathrm{Im}f^\mathrm{e}_z]=$$[\|\fep\| \sin\theta_{f^\mathrm{e}}\cos\varphi_{f^\mathrm{e}}, \|\fep\|\sin\theta_{f^\mathrm{e}}\sin\varphi_{f^\mathrm{e}}, \|\fep\|\cos\theta_{f^\mathrm{e}}]$ in the (b) 0 MF (c) 1 MF (d) 2 MF (e) 3 MF (f) 4 MF phases.
        The unit vector $\fep/\|\fep\|$ (upper figure) and the angle of the vector $\varphi_{f^\mathrm{e}},\theta_{f^\mathrm{e}}$ (lower figure) are plotted as a function of site $j$.
        We focus on near the edge of the system ($j\in[1,300]$).
        $\theta = \pi/4$, $\Delta_{\uparrow\uparrow}/t_1=0.18$, $\Delta_{\downarrow\downarrow}/t_1=1.8$, $h/t_1=2$, $\mu/t_1=1$, $L=5000$. 
        $\omega/t_1=E_G/2$: (a) $2.3\times 10^{-3}$, (b) $3.3\times 10^{-2}$, (c) $2.9\times 10^{-2}$, (d) $1.2\times 10^{-2}$, (e) $1.4\times 10^{-2}$.
        $\varphi'_{f^\mathrm{e}}$ and $\theta'_{f^\mathrm{e}}$: $\varphi_{f^\mathrm{e}}$ and $\theta_{f^\mathrm{e}}$ when $h_y\to -h_y$.
            Unlike the odd-frequency $\bm{\mathrm{f}}$ vector, the even-frequency $\bm{\mathrm{f}}$ vectors are not fixed in the same direction in the topological phases.
            Similar to the odd-frequency $\bm{\mathrm{f}}$ vector, inverting the sign of $h_y$ reverses the sign of $\mathrm{Im}f^\mathrm{e}_z$: $\theta_{f^\mathrm{e}}'=\pi - \theta_{f^e}$.
        \label{fig:direction-even-freq}
    }
\end{figure*}

To compare the odd-frequency $\bm{\mathrm{f}}$ vector to the even one, we calculate the spatial dependence of even one by using \eqs{(\ref{eq:odd-even-freq-vector}) and (\ref{eq:even-freq})}.
The spatial dependence of the even-frequency $\bm{\mathrm{f}}$ vector in 0--4 MF phases is shown in \figs{\ref{fig:spatial-even-freq}(a)--\ref{fig:spatial-even-freq}(e)}
.
In the trivial (0 MF) phase [\fig{\ref{fig:spatial-even-freq}(a)}], the pair amplitude oscillates throughout the space similar to the odd-frequency pair amplitude in \fig{\ref{fig:spatial-odd-freq}(a)}.
In the topological (1--4 MF) phases [\fig{\ref{fig:spatial-even-freq}(b)--(e)}], these values are constant near the bulk unlike the odd-frequency pair amplitude in \figs{\ref{fig:spatial-odd-freq}(b)--\ref{fig:spatial-odd-freq}(e)}.
Particularly, in the 3 and 4 MF phases, the amplitude oscillates near the edge as well as that of odd-frequency in \figs{\ref{fig:spatial-odd-freq}(d) and \ref{fig:spatial-odd-freq}(e)}.
The proportion of each component $|f_x^\mathrm{e}|$, $|f_y^\mathrm{e}|$, and $|f_z^\mathrm{e}|$ in 1--4MF [\figs{\ref{fig:spatial-even-freq}(b)--\ref{fig:spatial-even-freq}(e)}] are not equal unlike odd-frequency components [\figs{\ref{fig:spatial-odd-freq}(b)--\ref{fig:spatial-odd-freq}(e)}].

\begin{figure}[!htbp]
    \centering
    \includegraphics[width=0.49\textwidth]{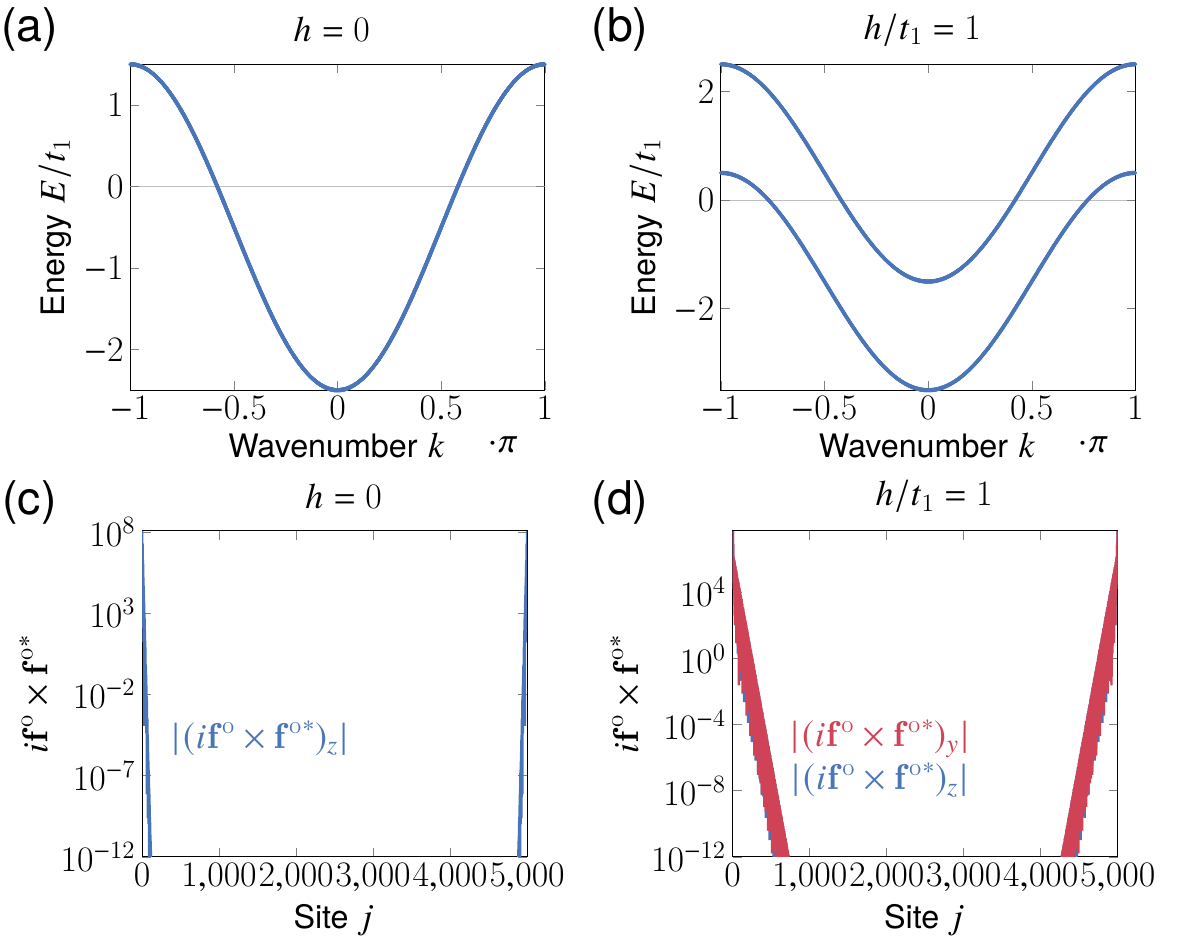}
    \caption{
        Energy bands of the Kitaev chain without pair potential for (a) $h=0$ and (b) $h/t_1=1$ in $\bm{\mathrm{k}}$ space.
        The spatial dependence of the spin of the odd-frequency Cooper pair states $\langle\sigma^\mathrm{o}\rangle\propto i \fo\times{\fo}^*$ for (c) $h=0$ and (d) $h/t_1=1$ in real space:
        $\omega/t_1=10^{-5}$, $L=5000$, $\Delta_{\uparrow\uparrow}/t_1=0.1$, $\Delta_{\downarrow\downarrow}=0$. $\theta = \pi/4$, $\mu/t_1=0.5$. 
            (c) The direction of $i \fo\times{\fo}^*$ inherits the spin direction of the pair potential ($i\bm{\mathrm{d}}\times\bm{\mathrm{d}}^\ast=[0,0,\Delta_{\uparrow\uparrow}^2/2]$).
            (d) The spin of the pair potential pointing in the $z$ direction is tilted by the magnetic fields in the $y$-$z$ plane.
        \label{fig:kitaev}
    }
\end{figure}

To investigate the proportion of the pair amplitude $|f_x^\mathrm{e}|$, $|f_y^\mathrm{e}|$, and $|f_z^\mathrm{e}|$ in detail, we have depicted the direction of the even-frequency $\bm{\mathrm{f}}$ vector [upper figures in \figs{\ref{fig:direction-even-freq}(b)--\ref{fig:direction-even-freq}(e)}], that is defined in \fig{\ref{fig:direction-even-freq}(a)}, and the angle $\theta_{f^\mathrm{e}}$ and $\varphi_{f^\mathrm{e}}$ [lower figures in \figs{\ref{fig:direction-even-freq}(b)--\ref{fig:direction-even-freq}(e)}] for $j\in[1,300]$ in the $L=5000$ sites system.
In the 0 MF phase, as shown in \fig{\ref{fig:direction-even-freq}(a)}, the even-frequency $\bm{\mathrm{f}}$ vector oscillates around the negative direction of the $x$ axis.
In the 1MF phase, the direction is almost along the negative direction of the $x$ axis without oscillation [\fig{\ref{fig:direction-even-freq}(c)}].
The even-frequency $\bm{\mathrm{f}}$ vectors in 2MF, 3MF, and 4MF phases are roughly oriented in the direction of between $-x$ and $-z$ axes, that of $y$ axis, that of between $-x$ and $y$ axes, respectively [\figs{\ref{fig:direction-even-freq}(d)--\ref{fig:direction-even-freq}(f)}, and also see \figs{\ref{fig:spatial-even-freq}(d)--\ref{fig:spatial-even-freq}(f)}].
The angle $\theta_{f^\mathrm{e}}'$ and $\varphi_{f^\mathrm{e}}'$ stand for $\theta_{f^\mathrm{e}}$ and $\varphi_{f^\mathrm{e}}$ when $h_y$ $\to$ $-h_y$, respectively.
While $\varphi_{f^\mathrm{e}}$ does not change ($\varphi_{f^\mathrm{e}}'=\varphi_{f^\mathrm{e}}$) with the respect to the sign change of $h_y$,
$\theta_{f^\mathrm{e}}$ change to $\pi-\theta_{f^\mathrm{e}}$ as shown in \figs{\ref{fig:direction-even-freq}(b)--\ref{fig:direction-even-freq}(f)}.

In this appendix, we have got two important results.
The even-frequency $\bm{\mathrm{f}}$ vector in each topological phase does not point in the same direction, unlike the odd-frequency one.
Similar to the odd-frequency $\bm{\mathrm{f}}$ vector, the sign inversion of $h_y$ inverts the sign of the z component of the even frequency $\bm{\mathrm{f}}$ vector.
These results support the discussion about the odd-frequency $\bm{\mathrm{f}}$ vector in \sect{\ref{sect:odd-freq}}.

\appsection{Spin structure of odd-frequency Cooper pairs in a Kitaev chain}
\label{sect:app_cross}
In \sect{\ref{sect:spatial}}, we have shown the spatial dependence of the spin states of the odd-frequency Cooper pairs $\langle\sigma^{\mathrm{o}}\rangle$ in a $p$-wave superconductor with magnetic fields.
The Kitaev chain, the simplest model expressing a spinless $p$-wave superconductor, is useful for investigating the effect of magnetic fields on the spin structure of the odd-frequency Cooper pairs. 
The Hamiltonian in \eqs{(\ref{eq:hamiltonian-real}), (\ref{eq:hamiltonian-k-one}), and (\ref{eq:hamiltonian-k-two})} can be rewritten as that in the Kitaev chain by replacing $\Delta_{\downarrow\downarrow}$, $t_2$, and $t_3\to0$. 

The plots of the energy dispersion in the bulk of the Kitaev chain without pair potential ($\Delta_{\uparrow\uparrow}=\Delta_{\downarrow\downarrow}=0$), shown in \figs{\ref{fig:kitaev}(a) and \ref{fig:kitaev}(b)}, are obtained by the diagonalization of the Hamiltonian in $\bm{\mathrm{k}}$ space that is expressed in \eq{\ref{eq:hamiltonian-k-two}}.
Without magnetic fields, the up and down spin states are degenerate to $E = - 2t_1\cos k -\mu$ [\fig{\ref{fig:kitaev}(a)}].
For $h/t_1=1$, these states split to $E = - 2t_1\cos k -\mu\pm h$  [\fig{\ref{fig:kitaev}(b)}].

The cross vector $i\bm{\mathrm{f}^{\mathrm{o} }}\times{\bm{\mathrm{f}^{\mathrm{o}} }}^*$ is proportional to the spin polarization of the odd-frequency Cooper pairs $\langle\sigma^{\mathrm{o}}\rangle$.
\headfig{\ref{fig:kitaev}(c)} [\headfig{\ref{fig:kitaev}(d)}] shows the cross vector $i\bm{\mathrm{f}^{\mathrm{o} }}\times{\bm{\mathrm{f}^{\mathrm{o}} }}^*$ in the Kitaev chain without [with] magnetic fields. 
The spin of pair potential $\langle\sigma\rangle \propto i\bm{\mathrm{d}}\times{\bm{\mathrm{d}^*}}=[0,0,\Delta_{\uparrow\uparrow}^2/2]$ points in the z direction.
Without the magnetic fields, only the $z$ component of $i\bm{\mathrm{f}^{\mathrm{o} }}\times{\bm{\mathrm{f}^{\mathrm{o}} }}^*$ exists since the spin state of the pair potential is inherited, as shown in \fig{\ref{fig:kitaev}(c)}.
When magnetic fields in the $y=z$ direction are applied, the cross vector has the $y$ component in addition to the z component [\fig{\ref{fig:kitaev}(d)}].
These results indicate that the spin of the odd-frequency Cooper pairs inherits that of the pair potential, and tends to point in the direction of the magnetic fields.

\bibliographystyle{apsrev4-2}
\bibliography{reference}

\end{document}